\documentclass[%
 reprint,
superscriptaddress,
 amsmath,amssymb,
 aps,
]{revtex4-2}

\usepackage{float}
\usepackage{graphicx}
\usepackage{chemformula}
\usepackage{braket}
\usepackage{soul}
\usepackage[compat=1.1.0]{tikz-feynman}
\usetikzlibrary{decorations.markings}
\usepackage{xcolor}
\tikzfeynmanset{warn luatex=false}
\definecolor{myviolet}{HTML}{660066}
\colorlet{myvioletlight}{myviolet!87!white}

\definecolor{myorange}{HTML}{FF6600}
\colorlet{myorangelight}{myorange!85!white}

\begin{document}

\begin{widetext}
\begin{abstract}
\noindent
The coupling between electronic excitations and collective bosonic modes is fundamental to the emergence of high-temperature superconductivity in cuprates. Despite extensive effort, conventional equilibrium and pump-probe optical spectroscopies still struggle to disentangle couplings to different bosonic modes when their energy scales overlap. Here we overcome this limitation using ultrafast two-dimensional electronic spectroscopy (2DES), which correlates coherent excitation and detection photon energies with femtosecond time resolution. Applied to optimally doped Bi$_2$Sr$_2$Ca$_{0.92}$Y$_{0.08}$Cu$_2$O$_{8+\delta}$, 2DES reveals a pronounced off-diagonal resonance arising from the ultrafast generation of non-thermal bosons with energy $\hbar\Omega_\mathbf{q}\simeq200$ meV. By comparing the measured spectra with a theoretical framework that explicitly includes the interaction between charge-transfer and magnetic excitations, we identify these bosons as paramagnons with momenta centered near $(\pi/2,\pi/2)$ and extending toward $(0,\pi)$ and $(\pi,0)$. The resonance persists across a large range of temperatures and doping concentrations, demonstrating that high-energy paramagnons are ubiquitously and strongly coupled to electronic excitations throughout the cuprate phase diagram. Time-domain analysis constrains the build-up of the paramagnon population to $\lesssim 10$ fs, placing a lower bound $\lambda \gtrsim 0.7$ on the coupling strength. More broadly, our results establish 2DES as a powerful approach for disentangling mode-selective electron-boson interactions and addressing decoherence dynamics, thereby establishing a new avenue for investigating strongly correlated quantum materials. These findings also provide a direct framework for future time-resolved resonant inelastic X-ray scattering experiments aimed at tracking the ultrafast dynamics of magnetic excitations.

\end{abstract}
\end{widetext}
\title{Ultrafast Two-Dimensional Spectroscopy Uncovers Ubiquitous Electron-Paramagnon Coupling in Cuprate Superconductors}

\author{Francesco Proietto}
\thanks{These authors contributed equally to this work.}
\affiliation{Department of Mathematics and Physics, Università Cattolica del Sacro Cuore, Brescia I-25133, Italy}
\affiliation{ILAMP (Interdisciplinary Laboratories for Advanced Materials Physics), Università Cattolica del Sacro Cuore, Brescia I-25133, Italy}
\affiliation{Department of Physics and Astronomy, KU Leuven, B-3001 Leuven, Belgium}
\author{Alessandra Milloch}
\thanks{These authors contributed equally to this work.}
\affiliation{Department of Mathematics and Physics, Università Cattolica del Sacro Cuore, Brescia I-25133, Italy}
\affiliation{ILAMP (Interdisciplinary Laboratories for Advanced Materials Physics), Università Cattolica del Sacro Cuore, Brescia I-25133, Italy}
\author{Paolo Franceschini}
\affiliation{CNR-INO (National Institute of Optics), via Branze 45, 25123 Brescia, Italy}
\affiliation{Department of Information Engineering, University of Brescia, Brescia I-25123, Italy}
\author{Mohammadjavad Azarm}
\affiliation{Department of Mathematics and Physics, Università Cattolica del Sacro Cuore, Brescia I-25133, Italy}
\affiliation{ILAMP (Interdisciplinary Laboratories for Advanced Materials Physics), Università Cattolica del Sacro Cuore, Brescia I-25133, Italy}
\affiliation{Department of Physics and Astronomy, KU Leuven, B-3001 Leuven, Belgium}
\author{Niccolò Sellati}
\affiliation{Department of Physics, ``Sapienza'' University of Rome, P.le A.\ Moro 5, 00185 Rome, Italy}
\author{Rishabh Mishra}
\affiliation{Optical Sciences Centre, Swinburne University of Technology, Melbourne, Australia}
\affiliation{ARC Centre of Excellence in Future Low-Energy Electronics Technologies, Swinburne University of Technology, Hawthorn, 3122, Victoria, Australia}
\affiliation{University of New South Wales, Sydney, Australia}
\author{Peter C. Moen}
\affiliation{Quantum Matter Institute, University of British Columbia, Vancouver, British Columbia, V6T 1Z4, Canada}
\affiliation{Department of Physics \& Astronomy, University of British Columbia, Vancouver, British Columbia, V6T 1Z1, Canada}
\author{Steef Smit}
\affiliation{Quantum Matter Institute, University of British Columbia, Vancouver, British Columbia, V6T 1Z4, Canada}
\affiliation{Department of Physics \& Astronomy, University of British Columbia, Vancouver, British Columbia, V6T 1Z1, Canada}
\author{Martin Bluschke}
\affiliation{Quantum Matter Institute, University of British Columbia, Vancouver, British Columbia, V6T 1Z4, Canada}
\affiliation{Department of Physics \& Astronomy, University of British Columbia, Vancouver, British Columbia, V6T 1Z1, Canada}
\author{Martin Greven}
\affiliation{School of Physics and Astronomy, University of Minnesota, Minneapolis, MN 55455, USA}
\author{Hiroshi Eisaki}
\affiliation{Core Electronics Technology Research Institute, National Institute of Advanced Industrial Science and Technology (AIST), 1-1-1 Umezono, Tsukuba, Ibaraki 305-8568, Japan}
\author{Marta Zonno}
\affiliation{Canadian Light Source, Inc., 44 Innovation Boulevard, Saskatoon, SK, Canada S7N 2V3}
\author{Sergey A. Gorovikov}
\affiliation{Canadian Light Source, Inc., 44 Innovation Boulevard, Saskatoon, SK, Canada S7N 2V3}
\author{Pinder Dosanjh}
\affiliation{Quantum Matter Institute, University of British Columbia, Vancouver, British Columbia, V6T 1Z4, Canada}
\affiliation{Department of Physics \& Astronomy, University of British Columbia, Vancouver, British Columbia, V6T 1Z1, Canada}
\author{Stefania Pagliara}
\affiliation{Department of Mathematics and Physics, Università Cattolica del Sacro Cuore, Brescia I-25133, Italy}
\affiliation{ILAMP (Interdisciplinary Laboratories for Advanced Materials Physics), Università Cattolica del Sacro Cuore, Brescia I-25133, Italy}
\author{Gabriele Ferrini}
\affiliation{Department of Mathematics and Physics, Università Cattolica del Sacro Cuore, Brescia I-25133, Italy}
\affiliation{ILAMP (Interdisciplinary Laboratories for Advanced Materials Physics), Università Cattolica del Sacro Cuore, Brescia I-25133, Italy}
\author{Fabio Boschini}
\affiliation{Advanced Laser Light Source, Institut National de la Recherche Scientifique, Varennes, QC J3X 1P7, Canada}
\author{Lara Benfatto}
\affiliation{Department of Physics, ``Sapienza'' University of Rome, P.le A.\ Moro 5, 00185 Rome, Italy}
\author{Giacomo Ghiringhelli}
\affiliation{Dipartimento di Fisica, Politecnico di Milano, piazza Leonardo da Vinci 32, I-20133
Milano, Italy}
\affiliation{CNR-SPIN, Dipartimento di Fisica, Politecnico di Milano, I-20133 Milano, Italy}
\author{Fulvio Parmigiani}
\affiliation{Elettra - Sincrotrone Trieste S.C.p.A., Trieste, Italy}
\affiliation{Dipartimento di Fisica, Università degli Studi di Trieste, Trieste, Italy}
\author{Jeffrey A. Davis}
\affiliation{Optical Sciences Centre, Swinburne University of Technology, Melbourne, Australia}
\affiliation{ARC Centre of Excellence in Future Low-Energy Electronics Technologies, Swinburne University of Technology, Hawthorn, 3122, Victoria, Australia}
\author{Andrea Damascelli}
\affiliation{Quantum Matter Institute, University of British Columbia, Vancouver, British Columbia, V6T 1Z4, Canada}
\affiliation{Department of Physics \& Astronomy, University of British Columbia, Vancouver, British Columbia, V6T 1Z1, Canada}
\author{Claudio Giannetti}
\thanks{claudio.giannetti@unicatt.it}
\affiliation{Department of Mathematics and Physics, Università Cattolica del Sacro Cuore, Brescia I-25133, Italy}
\affiliation{ILAMP (Interdisciplinary Laboratories for Advanced Materials Physics), Università Cattolica del Sacro Cuore, Brescia I-25133, Italy}
\affiliation{CNR-INO (National Institute of Optics), via Branze 45, 25123 Brescia, Italy}

\maketitle
The coupling of electronic quasiparticles to bosonic fluctuations is a key ingredient for explaining and ultimately controlling high-temperature superconductivity \cite{Keimer2015,Giannetti2016_NC,Sobota2021}. Unlike conventional superconductors, where the Bardeen-Cooper-Schrieffer (BCS) framework attributes Cooper pairing to phonon exchange, unconventional superconductors cannot be explained within a purely phonon-mediated picture \cite{Bardeen1957, Keimer2015, Scalapino2012}. In hole-doped copper oxides, the unconventional superconductors with the highest critical temperatures, the elusive pairing “glue” has been variously ascribed to lattice vibrations, spin fluctuations, or more exotic collective modes emerging from strong electronic correlations \cite{Scalapino2012,Keimer2015}.
Evidence for these interactions comes from a broad range of experimental probes. Angle-resolved photoemission spectroscopy (ARPES) has revealed dispersion anomalies (“kinks”) associated with bosonic modes \cite{Lanzara2001,Zhou2003,Damascelli2003}; optical and Raman spectroscopies have identified collective excitations in the charge and lattice channels \cite{Devereaux2007}; resonant inelastic X-ray scattering (RIXS) has probed high-energy spin and orbital excitations \cite{Minola2015}; and neutron scattering has mapped magnetic resonance modes in the superconducting state \cite{Eschrig2006}. Yet a fundamental limitation of equilibrium spectroscopies is the inability to disentangle electronic couplings to different bosonic channels when these occur on similar energy scales.

The emergence of nonequilibrium spectroscopies offered a new route to address this problem. By driving the system out of equilibrium and tracking its subsequent relaxation, it became possible to dynamically disentangle the electronic, lattice, and spin channels \cite{Giannetti2016,Boschini2024}. In particular, pump-probe optical measurements have revealed ultrafast energy exchange between photoexcited electrons and bosons, likely of magnetic origin \cite{DalConte2012,DalConte2015}. However, the strength of this coupling has so far been inferred only indirectly from the dynamics of an effective bosonic temperature, which affects the Drude scattering rate over a broad frequency range. Moreover, the lack of a controlled correlation between pump and probe frequencies has so far prevented an unambiguous identification of the bosons involved in the ultrafast dynamics.

Two-dimensional electronic spectroscopy (2DES) overcomes these limitations by simultaneously resolving the coherent electronic response as a function of both excitation and detection photon energies, while retaining the ultrafast temporal resolution of pump-probe techniques. The resulting two-dimensional spectra encode coherent and incoherent correlations between the initial and final states, establishing a direct link between the injected energy and the energy at which the optical response is detected. This capability grants access to time- and energy-resolved interaction pathways that remain hidden in single-color or broadband pump-probe spectroscopies \cite{Mukamel1995}.

Here, we exploit 2DES to investigate electron-boson interactions in the prototypical cuprate Bi$_2$Sr$_2$Ca$_{0.92}$Y$_{0.08}$Cu$_2$O$_{8+\delta}$ (Y-Bi2212). Upon excitation with ultrashort pulses, we observe a distinct off-diagonal resonance arising from the ultrafast creation and annihilation of nonthermal bosons at energies of the order of 200 meV, with momenta centered near $(\pi/2,\pi/2)$, extending towards $(0,\pi)$ and $(\pi,0)$. These characteristics are consistent with strong coupling to paramagnons and appear ubiquitously across the cuprate phase diagram over a broad hole-doping concentration and temperature range.
\begin{figure}[]
\includegraphics[width=8cm]{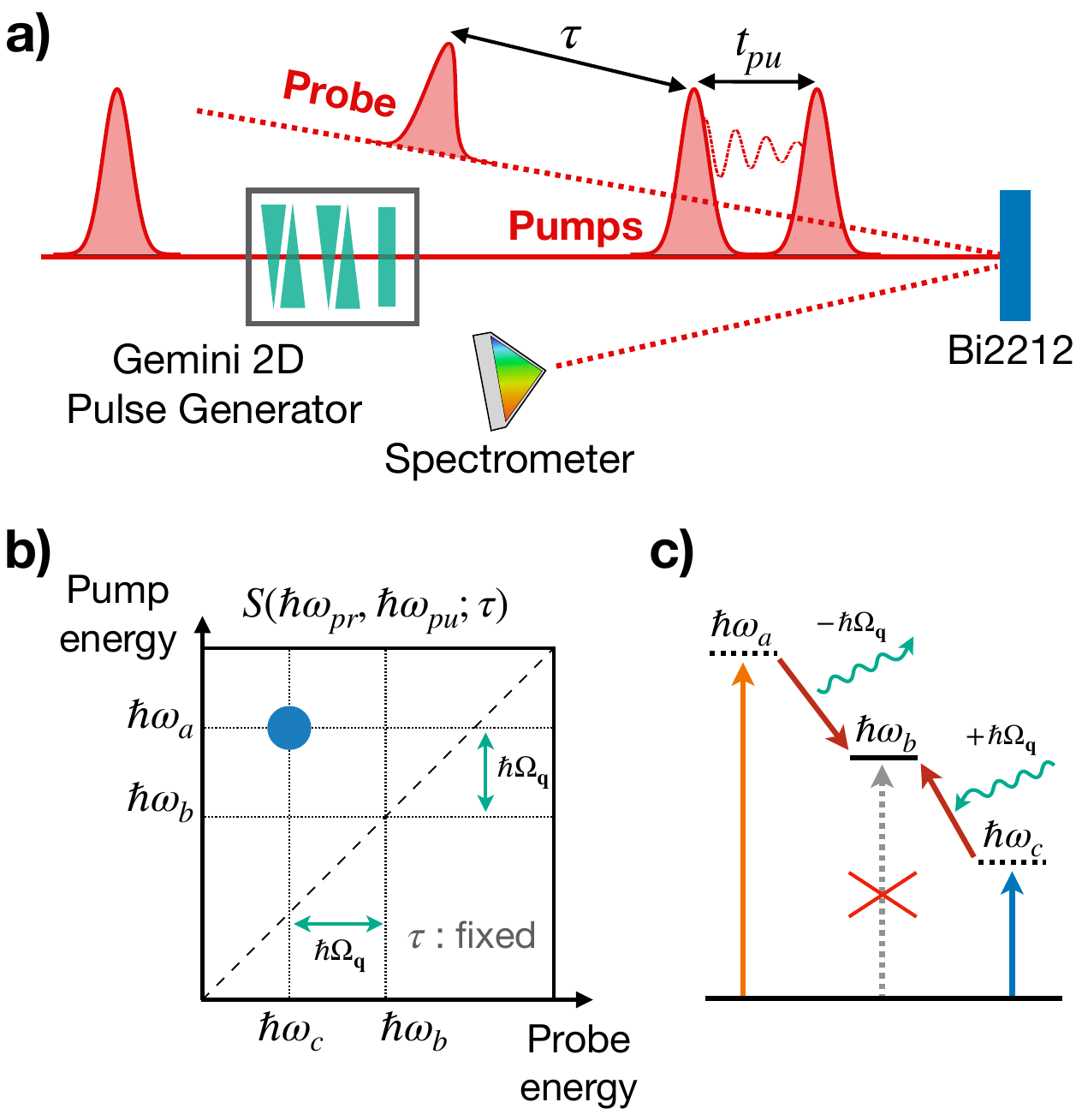}
\caption{(a) Schematic of the partially collinear 2DES setup. Two phase-locked broadband pump pulses, separated by a delay $t_{pu}$, excite the sample; after a waiting time $\tau$ a probe pulse measures the response. The reflected probe is spectrally dispersed by a spectrometer to define the detection axis $\hbar\omega_{pr}$. Pump replicas are produced by a common-path wedge interferometer to ensure phase stability and precise control of time delay $t_{pu}$ (see Supplementary Sec. S1). (b) Sketch of a 2D spectrum $S(\hbar\omega_{pr},\hbar\omega_{pu};\tau)$ at fixed $\tau$. The dashed diagonal marks $\hbar\omega_{pu}=\hbar\omega_{pr}$; an off-diagonal cross-peak ($\hbar\omega_c,\hbar\omega_a$) is indicated. (c) Illustration of a boson-mediated three-level scheme. First, pump photons drive a vertical transition to a virtual level at $\hbar\omega_a=\hbar\omega_b+\hbar\Omega_{\mathbf q}$, which relaxes to the real state $\hbar\omega_b$ by emitting a boson (energy $\hbar\Omega_{\mathbf q}$, momentum $\mathbf q$). This results in the creation of a transient non-thermal boson population peaked at $\hbar\Omega_{\mathbf q}$. Second, the probe promotes the system to a virtual intermediate state at $\hbar\omega_c=\hbar\omega_b-\hbar\Omega_{\mathbf q}$, and the transition is completed to the real state by absorbing a non-thermal boson of energy $\hbar\Omega_{\mathbf q}$. The corresponding cross-peak is observed at $(\hbar\omega_c,\hbar\omega_a)$ in $S(\hbar\omega_{pr},\hbar\omega_{pu};\tau)$ in panel (b).}
\label{fig:2des}
\end{figure}

2DES employs multiple light pulses to access the third-order electronic polarization $P^{(3)}(t)$ of a material {\cite{Mukamel1995,Mukamel2000,Hamm_Zanni_2011,Cundiff2013,Giannetti2016,Tollerud2017}}. In the simplest configuration, two phase-coherent pump pulses, separated by a delay $t_{pu}$, excite the sample (Fig. \ref{fig:2des}a) and a third probe pulse measures the optical response after a waiting time $\tau$. Fourier-transforming the signal with respect to $t_{pu}$ provides the excitation energy axis $\hbar\omega_{pu}$, while a spectrometer defines the detection axis $\hbar\omega_{pr}$. This yields a two-dimensional spectrum $S(\hbar\omega_{pr},\hbar\omega_{pu};\tau)$, schematically shown in Fig. \ref{fig:2des}b, which directly correlates excitation and detection electronic channels. The dashed diagonal, defined by $\hbar\omega_{pu}=\hbar\omega_{pr}$, corresponds to direct optical transitions that are excited and probed at the same photon energy. In contrast, off-diagonal peaks, for which $\hbar\omega_{pu}\neq\hbar\omega_{pr}$, indicate indirect pathways arising from coherent coupling or energy transfer between distinct electronic transitions. Compared to transient pump-probe spectroscopy, whose broadband spectrum is the 2DES spectrum integrated over the $\hbar \omega_{pu}$ axis (Projection slice theorem \cite{Gallagher99}), 2DES can resolve cross-peaks pathways from the diagonal peaks \cite{Hamm_Zanni_2011}.
More specifically, 2DES can reveal electronic excitations dressed by inelastic interaction with high-energy bosons, such as phonons or magnetic excitations \cite{Fuller2015,Cundiff2013}. Consider a high-energy state at $\hbar\omega_{b}$ that cannot be directly excited because of energy-momentum conservation.
Figure \ref{fig:2des}c shows a schematic boson-mediated process that can lead to off-diagonal signals in 2DES spectra. In a first step, the pump photon excites a vertical transition to a virtual state at energy $\hbar\omega_{a}=\hbar\omega_{b}+\hbar\Omega_\mathbf{q}$, which coherently couples to the real state $\hbar\omega_{b}$ by exchanging momentum $\mathbf{q}$ and emitting a boson of energy $\hbar\Omega_\mathbf{q}$. This process creates a transient coherent excitation of bosonic modes, which rapidly dephase, thus creating a non-thermal boson population with a distribution peaked at $\hbar\Omega_\mathbf{q}$. In the second step, the probe photon at energy $\hbar\omega_{c}=\hbar\omega_{b}-\hbar\Omega_\mathbf{q}$ is absorbed via annihilation of non-thermal bosons at energy $\hbar\Omega_\mathbf{q}$. This nonlinear process gives rise to an off-diagonal signal at $(\hbar\omega_c,\hbar\omega_a)$ in the $S(\hbar\omega_{pr},\hbar\omega_{pu};\tau)$ spectrum (see Fig. \ref{fig:2des}b), where the energy difference $\hbar\omega_a-\hbar\omega_c$ is approximately twice the boson energy $\hbar\Omega_\mathbf{q}$. 
\begin{figure*}[]
\includegraphics[width=14cm]{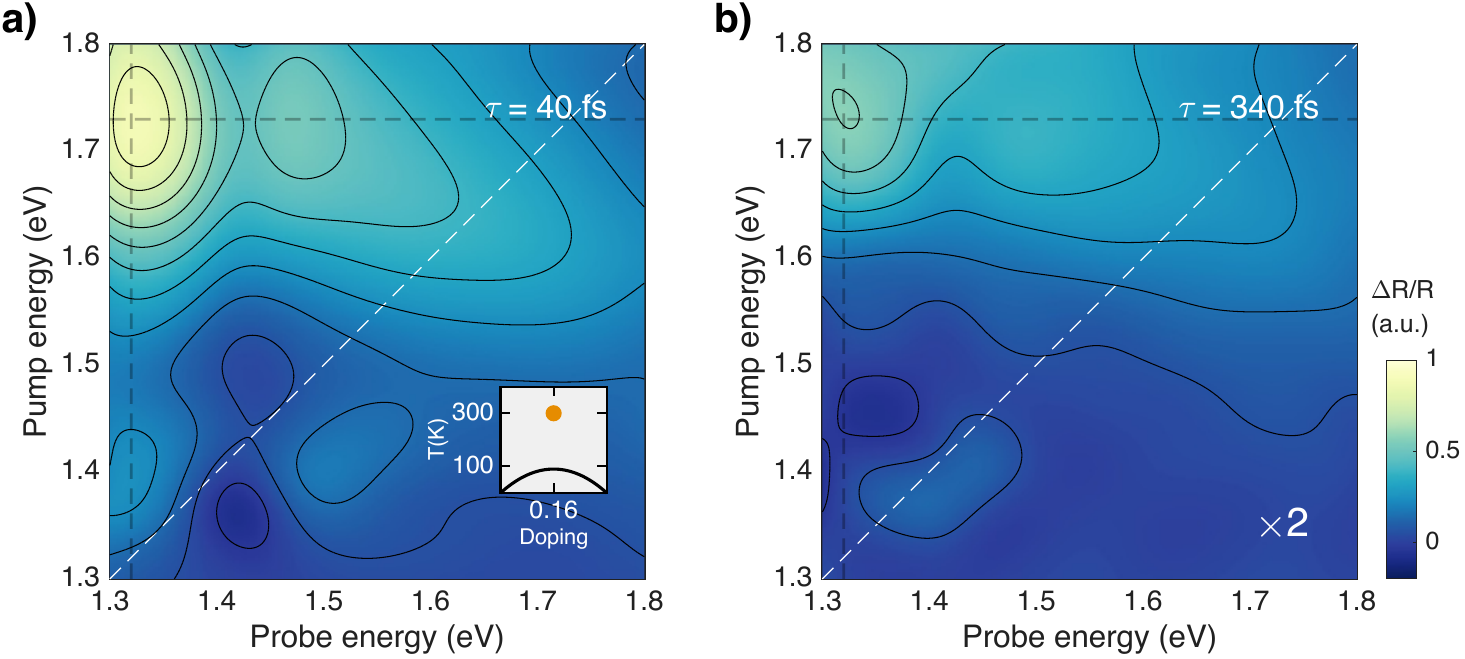}
\caption{(a) 2DES map of optimally doped Y-Bi2212 measured at $300$ K with a pump fluence of $400$ \textmu\text{J/cm}$^2$, recorded at a fixed delay time $\tau = 40$ fs. The color map shows the differential reflectivity $\Delta R/R$ as a function of pump photon energy $\hbar\omega_{pu}$ (vertical axis) and probe photon energy $\hbar\omega_{pr}$ (horizontal axis). The white dashed line marks the diagonal direction $\hbar\omega_{pu} = \hbar\omega_{pr}$. Black dashed lines indicate the position of the off-diagonal peak (see Supplementary Sec. S3). The inset shows the position of the measurement in the temperature-doping phase diagram. (b) 2DES map measured under the same experimental conditions as in panel (a) but at a longer delay time, $\tau = 340$ fs. The same color scale as in panel (a) is used; for clarity, the signal amplitude is multiplied by a factor of two. The off-diagonal resonance observed at $\tau = 40$ fs is preserved at long delay.}
\label{fig:2Ddatafull}
\end{figure*}

Bi-based cuprate superconductors are the ideal platform to investigate electron-boson coupling with 2DES. Y-Bi2212 presents a rich optical response in the near-infrared and visible range (see Supplementary Fig. S3a) \cite{Uchida1991}, and its connection to superconductivity has been extensively studied by both equilibrium \cite{Basov2005,Basov2011} and ultrafast techniques \cite{Giannetti2011,Giannetti2016}. As a general feature, the optical conductivity of undoped (and not superconducting) cuprates is characterized by a direct and intense charge-transfer peak at 2 eV \cite{Basov2005}. Upon doping, the strong absorption peak reminiscent of the direct charge-transfer (CT) transition moves to higher energies ($\gtrsim2.5-3$~eV), while the gap is filled with states corresponding to weak optical transitions, whose origin is still unclear \cite{Giannetti2011, Montanaro2024, Demedici2009}. These transitions involve both the O-2$p_{\pi}$ and O-2$p_{\sigma}$ bands \cite{Comanac2008}, which arise from the hybridization of the O-2$p_{x,y}$ orbitals perpendicular and parallel to the O-2$p$ ligand orbitals, respectively. 
Particular attention has been devoted to the study of a weak transition at $\hbar\omega\simeq 1.5$ eV, which is strongly affected by the onset of superconductivity \cite{Giannetti2011}. Considering that at this energy no direct vertical transition from the O-$2p_{\sigma,\pi}$ to the Cu-$3d_{x^2-y^2}$ is possible, as will be discussed in more detail in the following, the most plausible explanation is that this process involves an indirect and boson-assisted charge transfer transition between O-$2p_{\sigma,\pi}$ states at $1.5$~eV binding energy and Cu-$3d_{x^2-y^2}$ states close to the Fermi level.
We also note that indirect optical processes involving a rearrangement of the electronic distribution within the O-$2p_{\sigma,\pi}$ and Cu-$3d_{x^2-y^2}$ orbitals are inherently coupled to a perturbation of the local spin configuration \cite{Barantani2022} and are therefore associated with the emission or absorption of magnetic excitations.
To investigate the nature of these ultrafast electron-boson interactions, we performed 2DES experiments in the normal state of optimally doped Y-Bi2212 (hole doping concentration $p=0.16$, corresponding to $T_c=95$ K). As sketched in Fig. \ref{fig:2des}a, and detailed in Supplementary Sec. S1, we employed a partially collinear layout in reflection geometry, which directly measures the absorptive spectrum, corresponding to the sum of the real parts of the rephasing and non-rephasing signals \cite{Fuller2015}. Two phase-locked pump replicas are generated from a single broadband pulse by a common-path wedge interferometer (GEMINI-2D, NIREOS) \cite{Brida2012}, based on the Translating-Wedge-Based Identical Pulses eNcoding System (TWINS), which ensures ultra-high phase stability \cite{Brida2012} and provides a precisely tunable pump-pump delay $t_{pu}$. The two pump replicas propagate collinearly, while the probe arrives at a different angle. The reflected signal is then analyzed by a second common-path Fourier transform spectrometer (GEMINI, NIREOS) which defines the detection axis $\hbar\omega_{pr}$. In this configuration, the 2DES signal is the relative reflectivity variation resolved along the $\hbar\omega_{pr}$ and $\hbar\omega_{pu}$ axes, i.e. $S(\hbar\omega_{pr},\hbar\omega_{pu};\tau)=\Delta R/R(\hbar\omega_{pr},\hbar\omega_{pu};\tau)$. The experiment covers $1.3-1.8$ eV spectral range, with a temporal resolution of $\simeq40$ fs set by the pulse time duration (see Supplementary Sec. S1).

Figure \ref{fig:2Ddatafull}a shows the  2DES map acquired at $T=$300~K, $\tau=40$ fs and pump fluence $400$ \textmu\text{J/cm}$^2$, i.e., $S(\hbar\omega_{pr},\hbar\omega_{pu};\tau=40~\text{fs})$. 
Whereas the diagonal signal appears weak and extremely broad, thus suggesting the indirect nature of these optical transitions, a clear and pronounced off-diagonal resonance is observed at $(\hbar\omega_{pr},\hbar\omega_{pu})\simeq(1.32,1.73)$ eV. This off-diagonal signal, corresponding to a positive reflectivity variation, remains essentially unchanged when varying fluence (see Supplementary Sec. S4) and is compatible with a pump-induced absorption increase. In the Supplementary Information (Sec. S2) we model the transient optical properties starting from a parametrization of the equilibrium dielectric function. From the analysis of the vertical and horizontal line cuts of the 2DES map (see Supplementary Sec. S3), we extract the energy difference between the pump and probe excitations $\Delta E = \hbar\omega_{pu}-\hbar\omega_{pr}\simeq410$ meV, hereafter referred to as the correlation energy. Such a large correlation energy points to an indirect process mediated by emission and absorption of bosons with energy $\hbar\Omega_\mathbf{q}=\Delta E/2\simeq200$ meV. This energy lies well above the phonon spectrum of cuprates, which extends only up to $\simeq90$ meV \cite{Norman2006,Basov2005,DalConte2012}, while it is, instead, consistent with the energy scale of paramagnon excitations, i.e., dispersive spin fluctuations in the \ch{CuO2} planes observed throughout the phase diagram of doped cuprates by RIXS and neutron scattering \cite{LeTacon2011,Dean2013,Minola2015,Peng2018}. This energy scale is also consistent with the magnetic correlations emerging from inhomogeneous Mott-localization, as identified across several cuprate families via ARPES, tunneling, and optical spectroscopy \cite{Pelc2019,Pelc2020}.

The off-diagonal structure persists at longer waiting times, as shown in Fig. \ref{fig:2Ddatafull}b for the case $\tau = 340$ fs. This excludes any possible coherent artifact, which should be confined to the pulse temporal overlap \cite{Fiore2025} (see Supplementary Sec. S5), and demonstrates that the generation of non-thermal bosons of magnetic origin with energy $\hbar\Omega_\mathbf{q}\simeq200$ meV is inherent to the excitation process at all timescales.
\bigskip
\newline

\begin{figure}[]
\includegraphics[width=8cm]{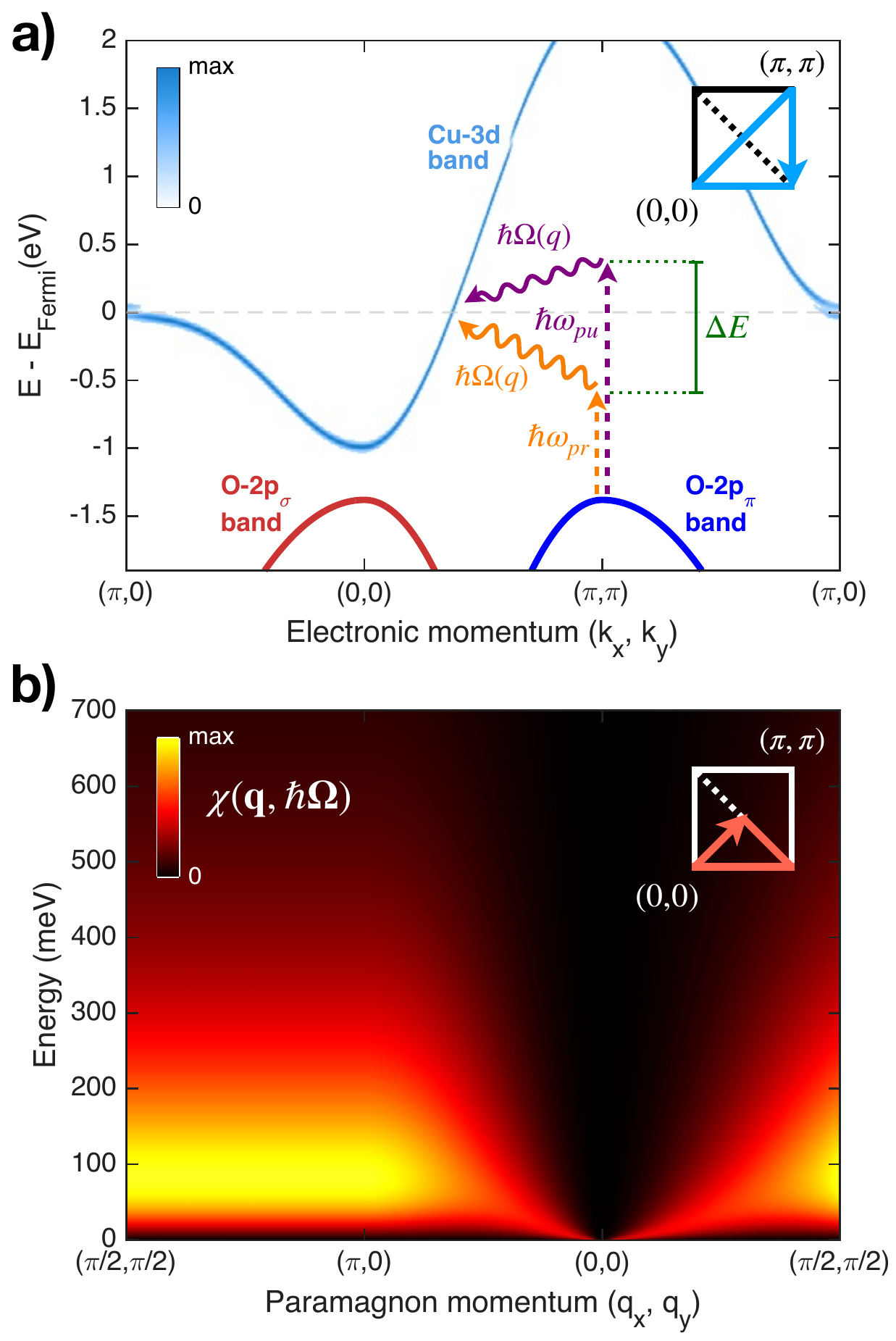}
\caption{Model of paramagnon-assisted charge-transfer pathways. (a) Schematic band structure illustrating the indirect optical transitions mediated by paramagnons. The O-$2p_{\sigma}$ (red) and O-$2p_{\pi}$ (blue) valence bands are shown together with the Cu-$3d_{x^2-y^2}$ conduction band (color scale), with energies referenced to the Fermi level $E_{Fermi}=\mu$. In the pump channel (purple arrows), a photon of energy $\hbar\omega_{pu}$ excites a vertical charge-transfer transition from the O-$2p$ band, followed by emission of a paramagnon with energy $\hbar\Omega_{\mathbf{q}}$. In the probe channel (orange arrows), a photon of energy $\hbar\omega_{pr}$ drives the transition via absorption of a previously generated paramagnon. The green bar indicates the resulting correlation energy $\Delta E = \hbar\omega_{pu} - \hbar\omega_{pr}$. The inset shows the high-symmetry path in the Brillouin zone along which the band dispersions are plotted. (b) Imaginary part of the spin susceptibility $\chi''(\mathbf{q},\hbar\Omega)$ used in the calculations, shown as a color map along a high-symmetry momentum path with $J=105$~meV and $\Gamma=300$~meV. The inset highlights the corresponding momentum direction in the Brillouin zone.}
\label{fig:model}
\end{figure}

\begin{figure*}[]
\includegraphics[width=12cm]{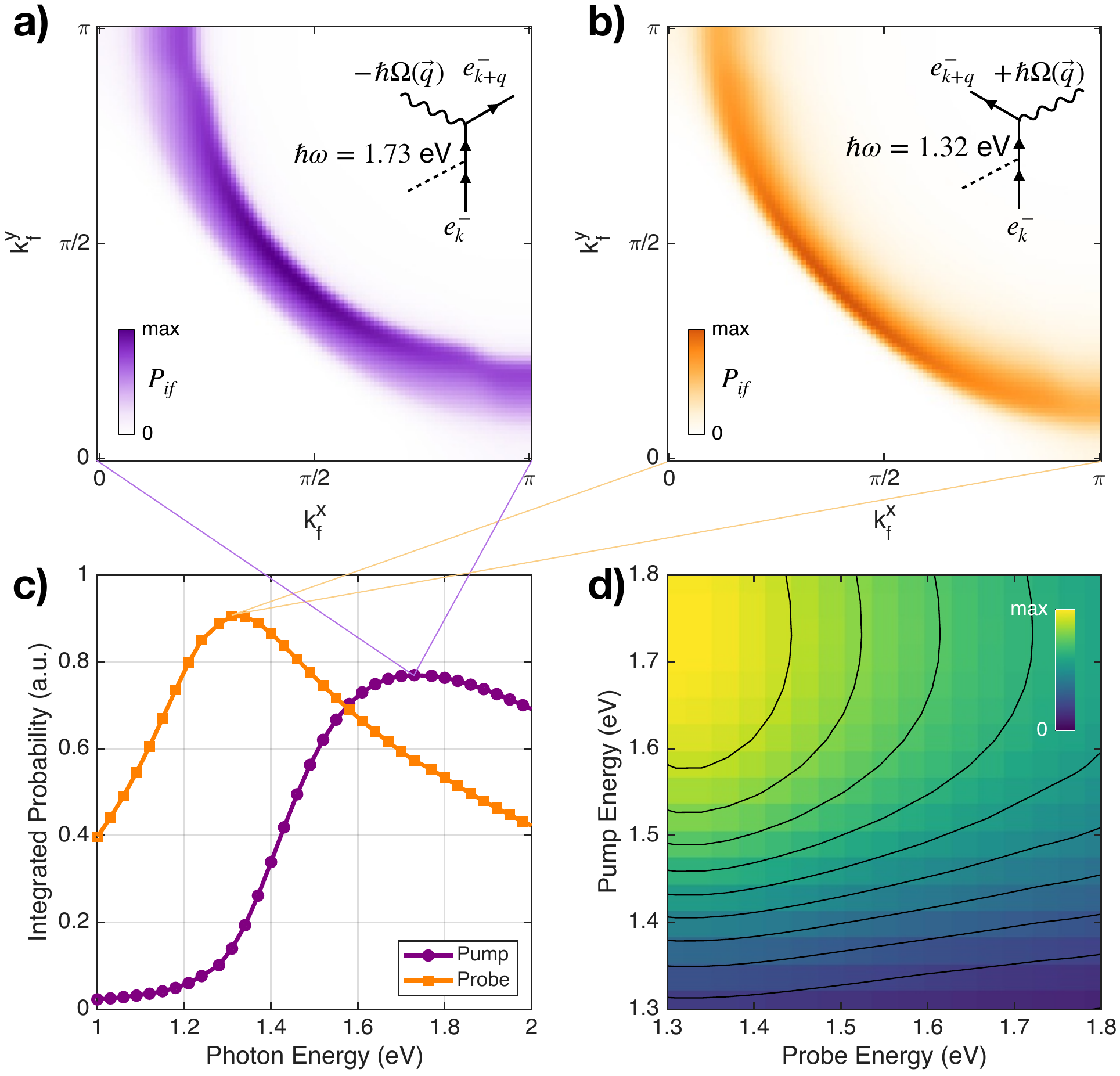}
\caption{(a) Momentum-resolved probability $P_{if}(\mathbf{k}_f)$ for the pump process at fixed photon energy $\hbar\omega=1.73$~eV, corresponding to excitation of an O-$2p_\pi$ electron followed by emission of a paramagnon with energy $\hbar\Omega_{\mathbf{q}}$, as sketched in the inset Feynman diagrams. 
(b) Momentum-resolved probability for the probe process at $\hbar\omega=1.32$~eV, involving absorption of a paramagnon previously generated by the pump, as sketched in the inset.
(c) Momentum-integrated transition probabilities for pump (purple) and probe (orange) processes as functions of photon energy, obtained by integrating, for each photon energy, the maps in panels (a) and (b) over the Brillouin zone.
(d) Calculated two-dimensional correlation map constructed as the product of the pump and probe energy dependences in panel (c), reproducing the off-diagonal resonance observed experimentally in 2DES.}
\label{fig:results}
\end{figure*}
To corroborate the interpretation of the 2DES data in terms of ultrafast coupling to paramagnons, we consider a minimal model based on the realistic electronic and magnetic dispersions and energy-momentum conservation. In Fig. \ref{fig:model}a we report the spectral functions of the O-$2p_{\sigma}$ and O-$2p_{\pi}$ bands (red and blue curves, respectively, extracted from Refs. \citenum{Ebrahimnejad2011,Cilento2018}) and of the Cu-3$d_{x^2-y^2}$ (colormap) bands. The latter is obtained from the analytical model in Ref. \citenum{Worm2024}, which provides an exactly solvable extension of the Hatsugai-Kohmoto model (HKM) \cite{Hatsugai1992}. In contrast to the original HKM, where the interaction is purely diagonal in momentum space, this formulation introduces a coupling between $\mathbf{k}$ and $\mathbf{k}+\mathbf{\Pi}$ with $\mathbf{\Pi}=(\pi,\pi)$, thus capturing the impact of short-range antiferromagnetic correlations and reproducing the band dispersion near the Fermi surface as well as the antinodal gap. Considering the O-$2p_{\pi}$ band, its  maximum lies at $E_{2p_{\pi}}\simeq-1.3$ eV for $\mathbf{k}_i=(\pi,\pi)$. Therefore, at the same momentum there are no available electronic states at the final energy $\hbar\omega-|E_{2p_{\pi}}|\lesssim0.1$ eV (see Fig. \ref{fig:model}a). The  O-$2p_{\pi} \rightarrow$ Cu-$3d_{x^2-y^2}$ charge-transfer excitation thus requires a process involving the transfer of both large momentum and energy  to reach available states near the Fermi energy. 
We describe the indirect optical transition within second-order perturbation theory. In the pump channel, absorption of a photon of energy $\hbar\omega_{pu}$ is accompanied by the emission of a paramagnon carrying energy $\hbar\Omega_{\mathbf q}$ and momentum $\mathbf q$, so that the electron reaches an available final state at energy $\hbar\omega_{pu}-\hbar\Omega_{\mathbf q}$ (purple path in Fig. \ref{fig:model}a). In the probe channel, the optical transition is instead assisted by the absorption of a paramagnon previously generated by the pump, allowing access to a final state at energy $\hbar\omega_{pr}+\hbar\Omega_{\mathbf q}$ (orange path in Fig. \ref{fig:model}a). The difference between these two processes corresponds to the correlation energy $\Delta E= 2\hbar\Omega_{\mathbf{q}}$ (green line in Fig. \ref{fig:model}a). For simplicity, in the following we label the initial O-$2p_{\pi}$ state as $\ket{i}$ and the final Cu-$3d_{x^2-y^2}$ state as $\ket{f}$. The probability of an indirect transition from $\ket{i}$ to $\ket{f}$, driven by the absorption of a photon at energy $\hbar\omega$ and assisted by the exchange of energy $\hbar\Omega_{\mathbf q}$ and momentum $\mathbf q$ with magnetic excitations, is \cite{Grosso2013, Fetter2003}:
\begin{align}
\label{eq:P}
P_{if}(\hbar\omega,\mathbf{k}_f)\propto \int\!\!&\int dE\,dE'\;f(E_i)[1-f(E_f)] \cdot \notag\\
& \cdot A_i(E - E_i(\mathbf{k}_i))\,A_f(E' - E_f(\mathbf{k}_f) + \hbar\omega)\cdot \notag \\ &\cdot\chi''(E' - E + \hbar\Omega_{\mathbf{q}}),
\end{align}
where the energy-conservation delta function has been decomposed into three terms and replaced by the spectral functions of the initial ($A_i(E,\mathbf{k}_i)$) and final ($A_f(E,\mathbf{k}_f)$) states, which account for lifetime broadening and band dispersion, together with the magnetic susceptibility $\chi''_{\mathbf q}(\hbar\Omega)$ which describes the bosons assisting the transition (see Supplementary Sec. S6A for more details). The Fermi-Dirac factors $f(E_i)$ and $[1-f(E_f)]$ account for the occupation of the initial and final electronic states at $T=300$ K.
Momentum conservation is enforced by imposing $\mathbf{k}_f=\mathbf{k}_i+\mathbf{q}$, where $\mathbf{k}_i$ and $\mathbf{k}_f$ denote the initial and final electron momenta. The initial-state spectral function $A_i=\eta/[(E-E_i)^2+\eta^2]$ is modeled as a Lorentzian with thermal broadening $\eta=k_B T=24$ meV, describing the O-$2p_{\pi}$ and O-$2p_{\sigma}$ bands \cite{Cilento2018}. For simplicity, we first consider the initial energy and momentum fixed at $E_i=-1.3$ eV and $\mathbf{k}_i= (\pi, \pi)$, corresponding to the maxima of the O-$2p_{\pi}$ states \cite{Ebrahimnejad2011,Cilento2018}. The final-state spectral function is obtained from the single-particle Green's function reported in Ref. \onlinecite{Worm2024}:
\begin{equation}
G_{\mathbf{k}}(E)= \frac{1-n_{\mathbf{k+\Pi}}}{E+\mu-\epsilon({\mathbf{k}})+i\eta}+ \frac{n_{\mathbf{k+\Pi}}}{E+\mu-\epsilon({\mathbf{k}})-\mathcal{V}+i\eta}, \end{equation}
using the relation $A_f(E)=-\frac{1}{\pi}\mathrm{Im}G_{\mathbf{k}}(E)$. Here, $n_{\mathbf{k+\Pi}}$ is the electronic occupation which determines the relative weight of the two poles in the Green's function, while the chemical potential $\mu=E_{Fermi}$ and the interaction parameter $\mathcal{V}$ are chosen to reproduce the Fermi surface measured via ARPES on the same sample (see Supplementary Sec. S6A and Fig. S7). The final-state dispersion $E_f(\mathbf{k})$ corresponds to $\epsilon(\mathbf{k})-\mu$, where $\epsilon(\mathbf{k})$ is the bare tight-binding dispersion with parameters $t=0.4$ eV, $t'=-0.08$ eV and $t''=0.04$ eV \cite{Norman2006, Damascelli2003,Comin2014} (see Supplementary Eq. S15).
The magnetic part is described by the imaginary part of the spin susceptibility $\chi(\mathbf q,\hbar\Omega)$, shown in Fig.~\ref{fig:model}b. We adopt the parametrization of Ref.~\onlinecite{LeTacon2011}, using the damping $\Gamma=300$~meV as measured by RIXS \cite{Peng2018} (see Supplementary Eq. S16-S17). The corresponding dispersion $\Omega_{\mathbf q}$ follows the standard spin-wave form for an antiferromagnetic square lattice. The superexchange coupling $J \simeq 105$ meV is fixed by requiring consistency with the experimentally observed peak separation in the 2DES correlation map, and is in line with literature values for cuprates \cite{LeTacon2011,Dean2013,Dean2013_PRL,Dahm2009}. Further analytical and computational details are provided in the Supplementary Sec. S6C. We note that the probability $P_{if}(\hbar\omega,\mathbf{k}_f)$ in Eq. \ref{eq:P} accounts only for energy and momentum conservation but does not consider the matrix elements of the optical transitions. A quantitative modeling of the absolute signal amplitude would, therefore, require a multiband evaluation of dipole matrix elements, which lies beyond the scope of the present minimal model. For a given photon energy $\hbar\omega$, the probability $P_{if}(\hbar\omega,\mathbf{k}_f)$ is peaked along two curves in momentum space, corresponding to available final states at energy $\hbar\omega-\hbar\Omega_{\mathbf{q}}$ for the paramagnon-emission channel and at energy $\hbar\omega+\hbar\Omega_{\mathbf{q}}$ for the paramagnon-absorption channel. For clarity, in the following, we present the two processes in separate panels (see Supplementary Sec. S6B for further details). Figure \ref{fig:results} reports the results for the O-$2p_{\pi}$ channel, obtained by fixing the initial momentum at $\mathbf{k}_i=(\pi,\pi)$. The corresponding calculation for the O-$2p_{\sigma}$ channel, with $\mathbf{k}_i=(0,0)$, is reported in Supplementary Fig. S8. Figure \ref{fig:results}a displays the calculated momentum-resolved probability $P_{if}(\mathbf{k})$ for $\hbar\omega=1.73$ eV, i.e., the photon energy at which the probability of a paramagnon-emission transition from O-$2p_\pi$ states to the Fermi level is maximized (see purple paths in Fig. \ref{fig:model}a). Figure \ref{fig:results}b displays the probability map for the probe channel, in which the optical transition is assisted by the absorption of a paramagnon previously generated during the pump process (see orange paths in Fig. \ref{fig:model}a). By integrating the momentum-resolved transition probability $P_{if}(\hbar\omega,\mathbf{k}_f)$ over the final momenta in the Brillouin zone, we obtain two distinct curves as a function of photon energy $\hbar\omega$, corresponding to the total probability of paramagnon emission (pump) and absorption (probe) processes, shown in Fig. \ref{fig:results}c. The resulting $P_{if}(\hbar\omega)$ exhibits maxima at $\simeq1.73$ eV and $1.32$ eV for the pump and probe channels, respectively, in agreement with the energy of the off-diagonal resonance observed in 2DES data. To enable a more direct comparison with the experiment, in Fig. \ref{fig:results}d we report the product of the two $P_{if}(\hbar\omega)$ curves corresponding to the paramagnon emission and absorption. The resulting correlation map quantitatively reproduces the position of the off-diagonal peak in the 2DES map in the ($\hbar\omega_{pr},\hbar\omega_{pu}$) plane, thereby supporting the interpretation in terms of strong coupling between indirect charge-transfer excitations at energy scales as high as $1.3-1.8$ eV and paramagnons with energy of the order of $\simeq200$ meV.

\begin{figure}[h]
\includegraphics[width=8cm]{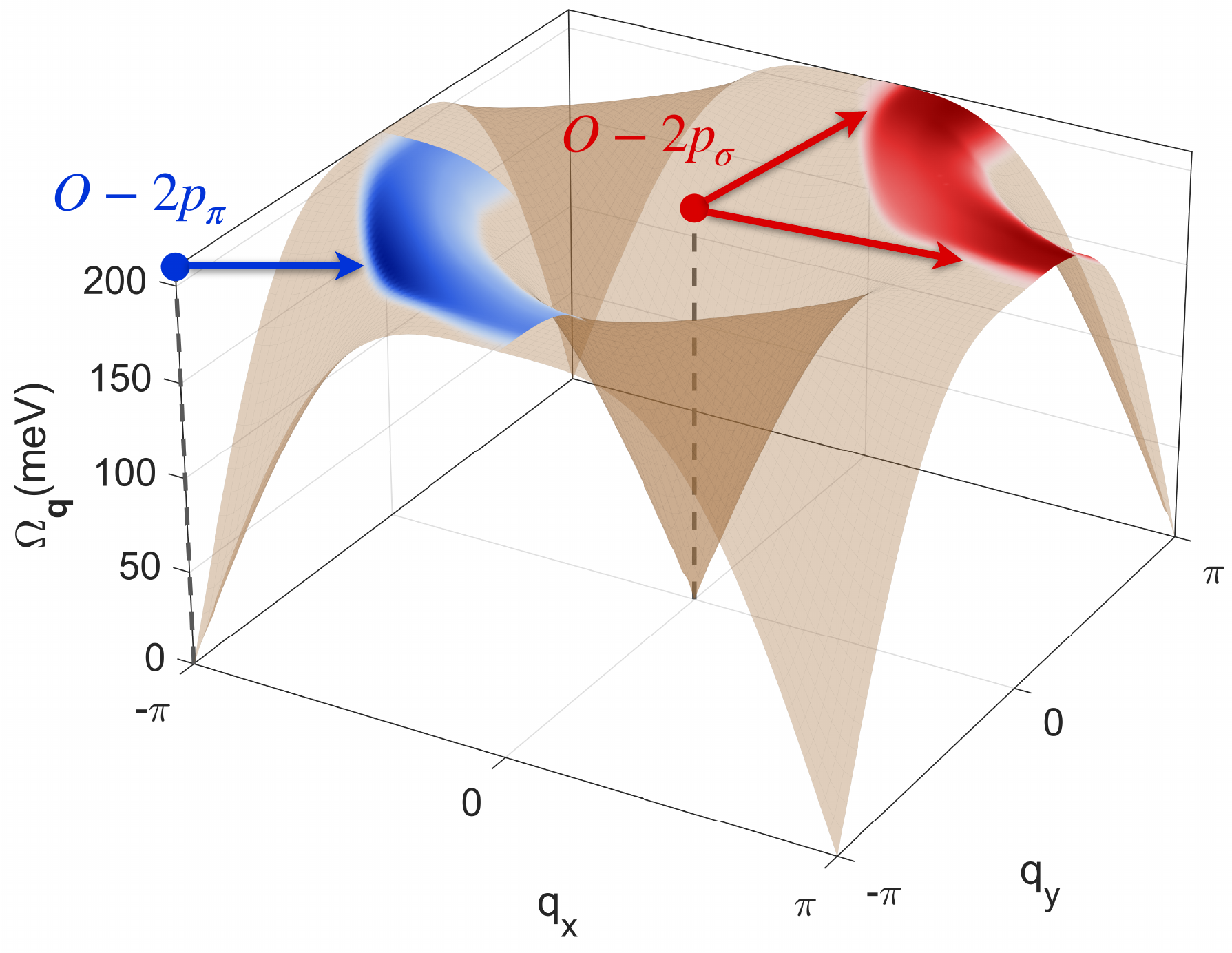}
\caption{Antiferromagnetic (paramagnon) dispersion and momentum-transfer phase space. The semi-transparent surface shows the model paramagnon dispersion $\hbar\Omega_{\mathbf{q}}$ over the full Brillouin zone. Overlaid is the probability $P_{if}$ for transitions mediated by paramagnon emission for incident photon energies between 1 and 2 eV. The darker blue (red) shaded areas highlight the set of momenta $\mathbf{q}$ contributing to this indirect process for the initial O-$2p_{\pi}$ [O-$2p_{\sigma}$] state. For visualization, the $(\pi,\pi)$-centered contribution is translated by a reciprocal-lattice vector and displayed around $(-\pi,-\pi)$ (BZ periodicity). Vertical lines mark the reference initial momentum points at $(-\pi,-\pi)$ (blue) and $(0,0)$ (red).}
\label{fig:dispersion}
\end{figure}

To visualize the energy-momentum transfer underlying the paramagnon-assisted charge-transfer pathway, Fig. \ref{fig:dispersion} shows the antiferromagnetic dispersion $\hbar\Omega_{\mathbf q}$ used in the calculations together with the set of momenta $\mathbf q$ contributing to the pump channel in the photon-energy window of Fig. \ref{fig:results}c. The blue [red] shaded region corresponds to momentum transfer from an initial O-$2p$ state at $(\pi,\pi)$ [$(0,0)$]. For clarity, the $(\pi,\pi)$-centered contribution is translated by a reciprocal-lattice vector and displayed around $(-\pi,-\pi)$. For the O-$2p_{\pi}$ initial state at $(\pi,\pi)$, the dominant contribution spans the portion of momentum space connecting the nodal region around $(\pi/2,\pi/2)$ to the antinodal regions near $(\pi,0)$ and $(0,\pi)$, while remaining concentrated near the upper part of the paramagnon band. The momentum map for the probe channel is reported in Supplementary Fig. S9. The corresponding calculation for the O-$2p_{\sigma}$ initial state at $(0,0)$, reported in Supplementary Fig. S8, does not reproduce the experimental correlation map with the same level of agreement.

An important question is whether the  ultrafast coupling with $\simeq 200$ meV paramagnons observed by 2DES at $T=300$ K is a specific characteristic of the high-temperature normal state or it affects the ultrafast dynamics also at lower temperatures and down to the superconducting state. To address this point, we performed similar 2DES measurements on optimally doped Y-Bi2212 ($T_c=95$ K) at $T=60$ K, i.e., deep in the superconducting phase, using a pump fluence of $400$~\textmu\text{J/cm}$^2$, which is large enough to non-thermally destroy the superconducting condensate \cite{Giannetti2016, Zonno2021}. The 2DES signal taken at $\tau=340$ fs (see Supplementary Sec. S7), i.e., before the recovery of the condensate within a few picoseconds, shows the same off-diagonal response as the $T=300$ K data, thus demonstrating that the coupling to paramagnons is a property of the normal state even at temperatures as low as those necessary for the onset of macroscopic superconductivity. By contrast, low-fluence measurements display a different behavior, whose detailed investigation is beyond the scope of the present work.

\begin{figure}[h]
\includegraphics[width=7cm]{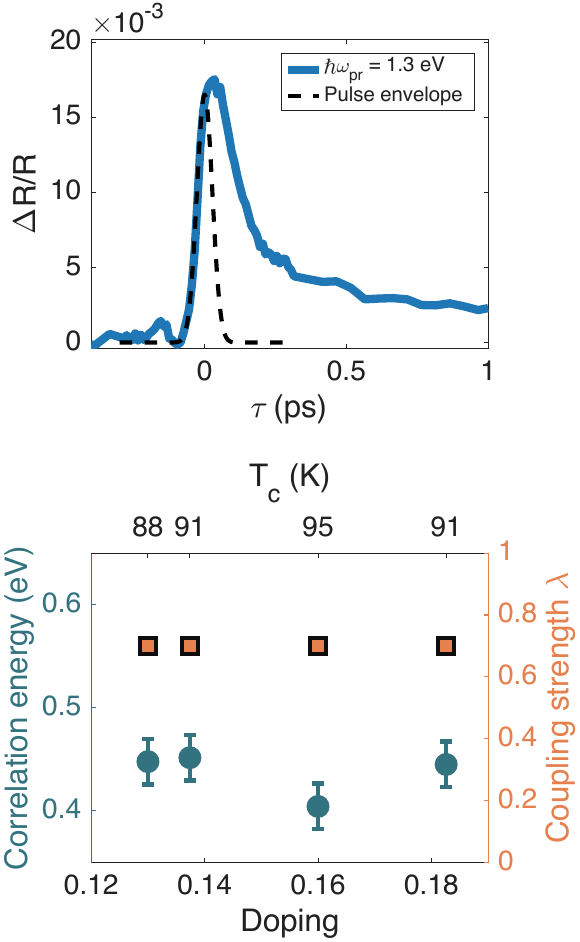}
\caption{(a) Time-resolved trace of the off-diagonal 2DES signal $S(\hbar\omega_{pr}=1.3~\mathrm{eV},\hbar\omega_{pu}=1.7~\mathrm{eV};\tau)$ measured on optimally doped Y-Bi2212 at $T=300$~K. The dashed line indicates the pump-probe pulse envelope. (b) Correlation energy $\Delta E=\hbar\omega_{pu}-\hbar\omega_{pr}$ (left axis, circles) and corresponding lower bound of the electron-boson coupling strength $\lambda$ (right axis, squares) extracted from the off-diagonal 2DES signal as a function of hole doping at $T=300$~K. The top axis shows the corresponding critical temperatures $T_c$(K). Error bars of 22 meV are estimated from peak position distribution in different measurements.}
\label{fig:dynamics}
\end{figure}

The simultaneous temporal and spectral resolution of 2DES allows us to probe the dynamics of the off-diagonal signal in the 2DES spectrum of Fig.~\ref{fig:2Ddatafull}, providing information on the timescale and strength of electron-paramagnon interactions. To reduce the total data acquisition time while still capturing the dynamics over a relevant temporal window, the probe spectrum has been filtered after the pump interaction with a 10 nm band-pass filter centered at 950 nm wavelength. The acquired data thus correspond to a vertical slice, centered at $\hbar\omega_{pr}=1.3$~eV, of the 2DES signal in Fig. \ref{fig:2Ddatafull}. In Fig. \ref{fig:dynamics}a we report the time trace at $\hbar\omega_{pr}=1.3$ eV, which corresponds to the signal $S(\hbar\omega_{pr}=1.3\text{ eV},\hbar\omega_{pu};\tau)$, integrated along the pump axis. We note that the off-diagonal feature appears essentially instantaneously, within the pump duration, and undergoes a multiexponential decay. This decay reflects the coupling to out-of-plane buckling and in-plane breathing Cu-O optical modes on a sub-200 fs timescale \cite{DalConte2012,DalConte2015}, followed by energy transfer to the remaining lattice modes on the picosecond timescale. Given the temporal resolution of the 2DES experiment, we can place an upper bound of $\simeq10$~fs for the build-up time of the hot boson population (see Supplementary Sec. S8 for details). Such a short timescale is consistent with the relaxation of photoexcited charges interacting with short-range antiferromagnetic excitations \cite{DalConte2015} and sets a lower bound on the electron-boson coupling strength. We estimate this bound using a conventional three-temperature model describing the energy exchange among the electronic excitations, the hot-boson population at $\hbar\Omega_{\mathbf q}\simeq200$ meV, and the lattice (see Supplementary Sec. S8).
In this model, the coupling to paramagnons is encoded in a bosonic spectral function $\alpha^2F(\omega)$ peaked at $200$ meV. The minimum coupling strength, which is compatible with a hot boson heating time of 10 fs is estimated to be $\lambda = 2 \int_0^\infty \frac{\alpha^2F(\omega)}{\omega} d\omega \simeq0.7$, thereby setting a lower bound for the electron-paramagnon coupling strength. This value is quite remarkable since it can, in principle, account for a very large critical temperature. Within a strong-coupling formalism, the critical temperature for $d$-wave pairing in a Fermi liquid is approximately given by $k_BT_c=0.83\times\hbar\Omega\times exp[-1.04(1+\lambda)/\lambda]$, assuming a vanishing pseudopotential \cite{Allen1975,Millis1988}. The combination of a strong selective coupling and a boson energy of $\hbar\Omega_{\mathbf{q}}\simeq200$ meV yields $T_c \simeq 150$ K, exceeding the actual $T_c$ of the material.
Finally, Fig.~\ref{fig:dynamics}b extends the analysis beyond the optimally doped compound to samples spanning the underdoped $(p=0.13)$ and overdoped $(p=0.18)$ regions of the phase diagram (see Supplementary Sec. S9). Remarkably, both the correlation energy $\Delta E$ and the corresponding lower bound on the coupling strength $\lambda$ remain essentially unchanged across the investigated doping range. This indicates that the ultrafast interaction with $\simeq200$ meV paramagnons persists broadly across the hole-doped cuprates phase-diagram.

In conclusion, we have introduced 2DES as a novel tool for investigating electron-boson coupling in the normal state of superconducting copper oxides. By combining temporal resolution with independent control of the excitation and detection energies, 2DES reveals a selective coupling between indirect charge-transfer O-$2p_{\pi}$ orbitals and Cu-$3d_{x^2-y^2}$ states at the Fermi level and paramagnons with energy $\hbar\Omega_{\mathbf q}\simeq 200$ meV. Although these charge excitations lie far above the energy scale of the superconducting gap, charge-transfer processes are intimately connected to superconductivity, as demonstrated by the well-established superconductivity-induced spectral-weight changes at energy scales as high as 1-2 eV \cite{Molegraaf2002,Boris2004,Basov2005,Basov2011,Giannetti2011,Giannetti2016}. The ultrastrong coupling between high-energy charge excitations and paramagnons therefore emerges as a key ingredient for understanding the microscopic origin of high-temperature superconductivity in cuprates. Our results constitute a benchmark for the next-generation of time-resolved RIXS experiments that should directly probe magnetic excitations and shed further light on the nature of the coupling between charge and magnetic excitations \cite{Chen2019,Mitrano2020,Hales2023,Mitrano2024,Jost2025}. More generally, this work shows that 2DES adds a crucial new dimension to time-resolved spectroscopies for disentangling charge and magnetic excitations in any family of correlated quantum materials, in which the spin, lattice, orbital, and electronic degrees of freedom are intrinsically intertwined.
\section*{Acknowledgments}
The authors thank G. Perosa for helpful discussion. Fr.P., A.M., M.A., and C.G. acknowledge financial support from MIUR through the PRIN 2020 (Prot. 2020JLZ52N-003) program and from the European Union - Next Generation EU through the MUR-PRIN2022 (Prot. 20228YCYY7) program. The work at the University of Minnesota was funded by the Department of Energy through the University of Minnesota Center for Quantum Materials under Award No. DE-SC0016371. This research was undertaken thanks in part to funding from the Max Planck–UBC–UTokyo Centre for Quantum Materials and the Canada First Research Excellence Fund (CFREF), Quantum Materials and Future Technologies. This project is also funded by the Natural Sciences and Engineering Research Council of Canada (NSERC), the Canada Foundation for Innovation (CFI); the Department of National Defence (DND); the British Columbia Knowledge Development Fund (BCKDF); the Canada Research Chairs (CRC) Program (A.D.); and the CIFAR Quantum Materials Program (A.D.). Use of the Canadian Light Source (Quantum Materials Spectroscopy Centre), a national research facility of the University of Saskatchewan, is supported by CFI, NSERC, the National Research Council (NRC), the Canadian Institutes of Health Research (CIHR), the Government of Saskatchewan and the University of Saskatchewan. S.S. acknowledges the support of the Netherlands Organization for Scientific Research (NWO 019.223EN.014, Rubicon 2022-3).
\bibliography{refs}

\end{document}


\title{Supplementary Information for \\ Ultrafast Two-Dimensional Spectroscopy Uncovers Ubiquitous Electron-Paramagnon Coupling in Cuprate Superconductors}

\author{Francesco Proietto}
\thanks{These authors contributed equally to this work.}
\affiliation{Department of Mathematics and Physics, Università Cattolica del Sacro Cuore, Brescia I-25133, Italy}
\affiliation{ILAMP (Interdisciplinary Laboratories for Advanced Materials Physics), Università Cattolica del Sacro Cuore, Brescia I-25133, Italy}
\affiliation{Department of Physics and Astronomy, KU Leuven, B-3001 Leuven, Belgium}
\author{Alessandra Milloch}
\thanks{These authors contributed equally to this work.}
\affiliation{Department of Mathematics and Physics, Università Cattolica del Sacro Cuore, Brescia I-25133, Italy}
\affiliation{ILAMP (Interdisciplinary Laboratories for Advanced Materials Physics)}
\author{Paolo Franceschini}
\affiliation{CNR-INO (National Institute of Optics), via Branze 45, 25123 Brescia, Italy}
\affiliation{Department of Information Engineering, University of Brescia, Brescia I-25123, Italy}
\author{Mohammadjavad Azarm}
\affiliation{Department of Mathematics and Physics, Università Cattolica del Sacro Cuore, Brescia I-25133, Italy}
\affiliation{ILAMP (Interdisciplinary Laboratories for Advanced Materials Physics), Università Cattolica del Sacro Cuore, Brescia I-25133, Italy}
\affiliation{Department of Physics and Astronomy, KU Leuven, B-3001 Leuven, Belgium}
\author{Niccolò Sellati}
\affiliation{Department of Physics, ``Sapienza'' University of Rome, P.le A.\ Moro 5, 00185 Rome, Italy}
\author{Rishabh Mishra}
\affiliation{Optical Sciences Centre, Swinburne University of Technology, Melbourne, Australia}
\affiliation{ARC Centre of Excellence in Future Low-Energy Electronics Technologies, Swinburne University of Technology, Hawthorn, 3122, Victoria, Australia}
\affiliation{University of New South Wales, Sydney, Australia}
\author{Peter C. Moen}
\affiliation{Quantum Matter Institute, University of British Columbia, Vancouver, British Columbia, V6T 1Z4, Canada}
\affiliation{Department of Physics \& Astronomy, University of British Columbia, Vancouver, British Columbia, V6T 1Z1, Canada}
\author{Steef Smit}
\affiliation{Quantum Matter Institute, University of British Columbia, Vancouver, British Columbia, V6T 1Z4, Canada}
\affiliation{Department of Physics \& Astronomy, University of British Columbia, Vancouver, British Columbia, V6T 1Z1, Canada}
\author{Martin Bluschke}
\affiliation{Quantum Matter Institute, University of British Columbia, Vancouver, British Columbia, V6T 1Z4, Canada}
\affiliation{Department of Physics \& Astronomy, University of British Columbia, Vancouver, British Columbia, V6T 1Z1, Canada}
\author{Martin Greven}
\affiliation{School of Physics and Astronomy, University of Minnesota, Minneapolis, MN 55455, USA}
\author{Hiroshi Eisaki}
\affiliation{Core Electronics Technology Research Institute, National Institute of Advanced Industrial Science and Technology (AIST), 1-1-1 Umezono, Tsukuba, Ibaraki 305-8568, Japan}
\author{Marta Zonno}
\affiliation{Canadian Light Source, Inc., 44 Innovation Boulevard, Saskatoon, SK, Canada S7N 2V3}
\author{Sergey A. Gorovikov}
\affiliation{Canadian Light Source, Inc., 44 Innovation Boulevard, Saskatoon, SK, Canada S7N 2V3}
\author{Pinder Dosanjh}
\affiliation{Quantum Matter Institute, University of British Columbia, Vancouver, British Columbia, V6T 1Z4, Canada}
\affiliation{Department of Physics \& Astronomy, University of British Columbia, Vancouver, British Columbia, V6T 1Z1, Canada}
\author{Stefania Pagliara}
\affiliation{Department of Mathematics and Physics, Università Cattolica del Sacro Cuore, Brescia I-25133, Italy}
\affiliation{ILAMP (Interdisciplinary Laboratories for Advanced Materials Physics), Università Cattolica del Sacro Cuore, Brescia I-25133, Italy}
\author{Gabriele Ferrini}
\affiliation{Department of Mathematics and Physics, Università Cattolica del Sacro Cuore, Brescia I-25133, Italy}
\affiliation{ILAMP (Interdisciplinary Laboratories for Advanced Materials Physics), Università Cattolica del Sacro Cuore, Brescia I-25133, Italy}
\author{Fabio Boschini}
\affiliation{Advanced Laser Light Source, Institut National de la Recherche Scientifique, Varennes, QC J3X 1P7, Canada}
\author{Lara Benfatto}
\affiliation{Department of Physics, ``Sapienza'' University of Rome, P.le A.\ Moro 5, 00185 Rome, Italy}
\author{Giacomo Ghiringhelli}
\affiliation{Dipartimento di Fisica, Politecnico di Milano, piazza Leonardo da Vinci 32, I-20133
Milano, Italy}
\affiliation{CNR-SPIN, Dipartimento di Fisica, Politecnico di Milano, I-20133 Milano, Italy}
\author{Fulvio Parmigiani}
\affiliation{Elettra - Sincrotrone Trieste S.C.p.A., Trieste, Italy}
\affiliation{Dipartimento di Fisica, Università degli Studi di Trieste, Trieste, Italy}
\author{Jeffrey A. Davis}
\affiliation{Optical Sciences Centre, Swinburne University of Technology, Melbourne, Australia}
\affiliation{ARC Centre of Excellence in Future Low-Energy Electronics Technologies, Swinburne University of Technology, Hawthorn, 3122, Victoria, Australia}
\author{Andrea Damascelli}
\affiliation{Quantum Matter Institute, University of British Columbia, Vancouver, British Columbia, V6T 1Z4, Canada}
\affiliation{Department of Physics \& Astronomy, University of British Columbia, Vancouver, British Columbia, V6T 1Z1, Canada}
\author{Claudio Giannetti}
\thanks{claudio.giannetti@unicatt.it}
\affiliation{Department of Mathematics and Physics, Università Cattolica del Sacro Cuore, Brescia I-25133, Italy}
\affiliation{ILAMP (Interdisciplinary Laboratories for Advanced Materials Physics), Università Cattolica del Sacro Cuore, Brescia I-25133, Italy}
\affiliation{CNR-INO (National Institute of Optics), via Branze 45, 25123 Brescia, Italy}

\maketitle
\section{Experimental setup}
\begin{figure}[t]
    \centering
    \includegraphics[width=0.98\linewidth]{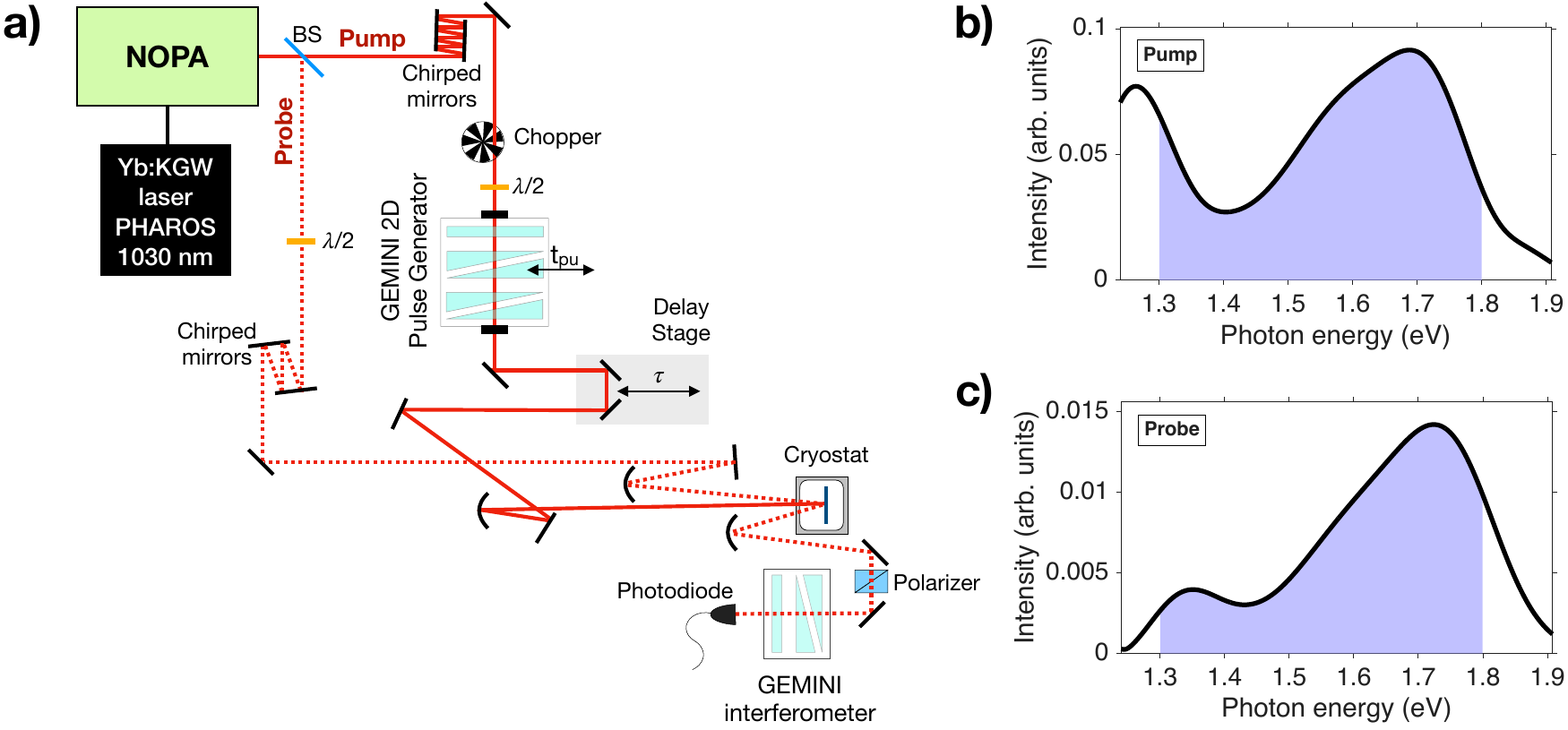}
    \caption{(a) Layout of the 2DES experimental setup. A broadband pulse generated by a NOPA is split into pump and probe beams; the pump is chopped and sent through the GEMINI-2D Pulse Generator to generate two phase-locked pump pulses with tunable coherence delay $t_{pu}$, while a motorized delay line sets the waiting time $\tau$ with respect to the probe. After interaction with the sample in the cryostat, the probe is analyzed with a Fourier-transform interferometer (GEMINI interferometer) and detected on a photodiode. (b) Measured spectra of the pump (top panel) and probe (bottom panel) beams. The shaded violet area indicates the spectral window considered for the analysis of the 2D spectra.}
    \label{fig:setup}
\end{figure}

The optical setup (Fig.~\ref{fig:setup}a) is designed to implement broadband two-dimensional electronic spectroscopy (2DES) in a partially collinear geometry. Both pump and probe pulses are generated by a home-built noncollinear optical parametric amplifier (NOPA) pumped by an Yb:KGW laser system (Pharos, Light Conversion), delivering pulses at $1030$~nm with a maximum repetition rate of $200$~kHz. The repetition rate is adjustable via a pulse picker; in the measurements presented here, $40$~kHz was used for high-fluence conditions (400 \textmu\text{J/cm}$^2$), while higher repetition rates were employed for lower fluences (up to 200 kHz for fluence of 40 \textmu\text{J/cm}$^2$). Pump and probe beams are obtained by splitting the NOPA output with a $10{:}90$ beam splitter. Figure~\ref{fig:setup}b shows the measured spectra of the pump and probe beams, with the violet shaded region indicating the spectral window used in the data analysis.
In the pump arm, the beam is modulated by a mechanical chopper to enable lock-in detection of the pump-induced reflectivity variation. The modulated pump is then directed into a common-path birefringent interferometer (GEMINI-2D, NIREOS), based on the Translating-Wedge-Based Identical Pulses eNcoding System (TWINS) scheme~\cite{Brida2012}, which generates a pair of phase-locked excitation pulses separated by a variable delay $t_{pu}$ (coherence time). A mechanical delay stage in the pump path introduces an additional delay $\tau$ (waiting time) by varying the optical path length of the pump relative to the probe.
Pump and probe pulses, orthogonally polarized to minimize pump scattering into the detection channel, are then focused onto the sample by reflective optics to 160$\times$160 \textmu m$^2$ and 80$\times$80 \textmu m$^2$ spot sizes, respectively. The pump beam impinges on the sample at normal incidence, while the probe beam is incident at an angle of approximately $10^\circ$. The samples are mounted in a closed-cycle helium cryostat.
Dispersion introduced by transmissive optical elements (including the GEMINI-2D interferometer, beam splitter, wave plates, and cryostat window) is compensated by two independent chirped-mirror compressors in the pump and probe paths, ensuring minimal residual chirp at the sample position. The 2DES signal at each $(t_{pu}, \tau)$ is collected in the reflected probe direction and sent to a Fourier-transform interferometer (GEMINI, NIREOS), which provides the transient reflectivity signal $\Delta R/R$ spectrally resolved along the probe photon-energy axis. A Fourier transform with respect to $t_{pu}$ yields the spectral resolution along the pump photon-energy axis. Finally, the measured two-dimensional signal is normalized with respect to both the pump and probe spectra.

\begin{figure*}[h]
\includegraphics[width=13.5cm]{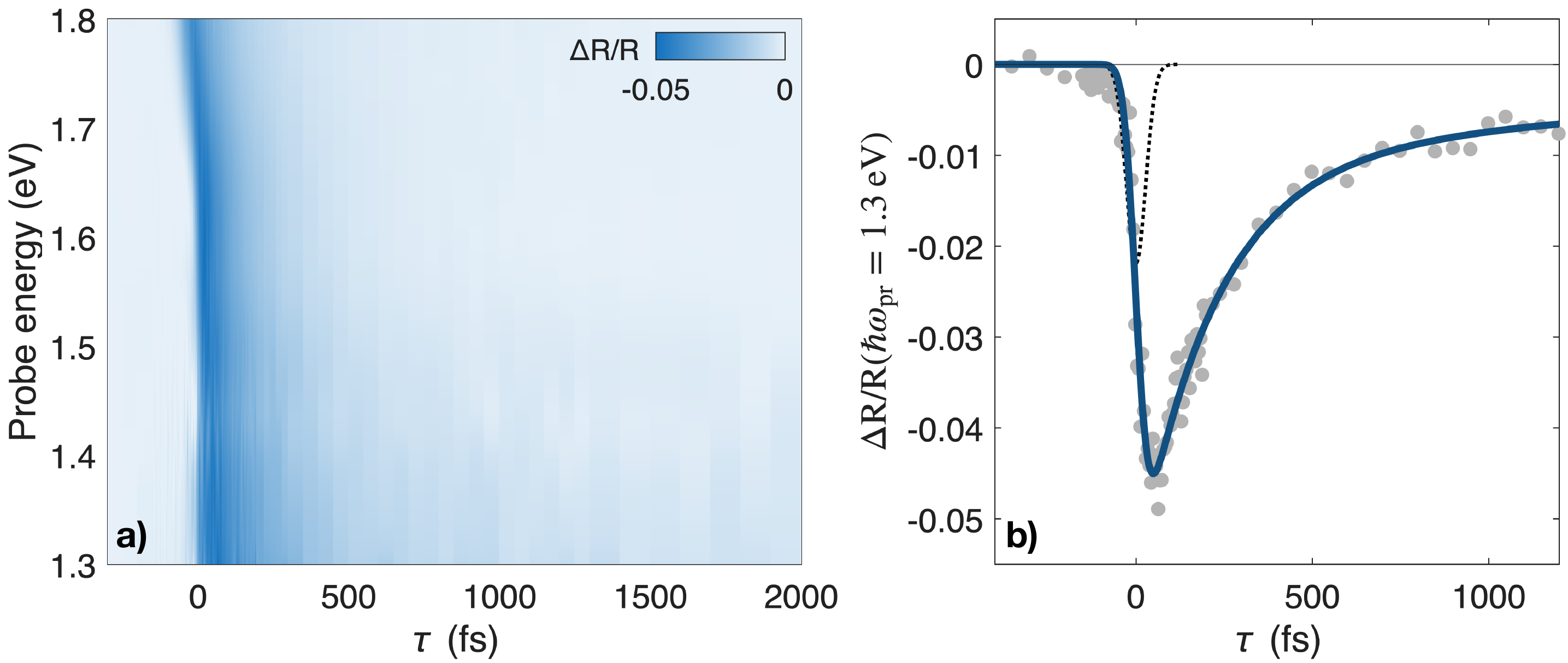}
\caption{(a) Transient reflectivity pump–probe measurement of a graphite sample. (b) Line profile at 1.3~eV and corresponding fit of the dynamics, yielding a pump–probe cross-correlation of 65~fs (black dotted line).}
\label{fig: time duration}
\end{figure*}

The temporal resolution of the experimental setup was characterized using a pump--probe measurement on Highly Oriented Pyrolytic Graphite (HOPG), which exhibits an essentially instantaneous response. The broadband probe measurement (Fig.~\ref{fig: time duration}a) shows no significant wavelength-dependent delays, indicating minimal residual chirp and proper optimization of the pulse compressors.

Since all transient dynamics discussed in the manuscript are analyzed at a probe photon energy of 1.3~eV (see Section~\ref{sec: coupling strength estimate}), we determine the effective temporal resolution by fitting the graphite pump-probe trace at this photon energy. The data are well described by a double-exponential decay convoluted with a Gaussian function accounting for the finite pulse duration. The Gaussian width is left as a free parameter, and the fit (see Fig.~\ref{fig: time duration}b) yields a full width at half maximum of  $65\pm6$~fs, which represents the pump--probe cross-correlation. Assuming similar pulse durations for the pump and probe, this corresponds to an individual pulse duration of approximately 40~fs.  The extracted pulse duration is consistent with previous FROG (Frequency Resolved Optical Gating) measurements performed on the same apparatus under similar operating conditions \cite{Milloch2026, Azarm2026}.

\clearpage
\newpage
\section{Optical conductivity and sign of the 2DES signal}

\begin{figure*}[h]
\includegraphics[width=12cm]{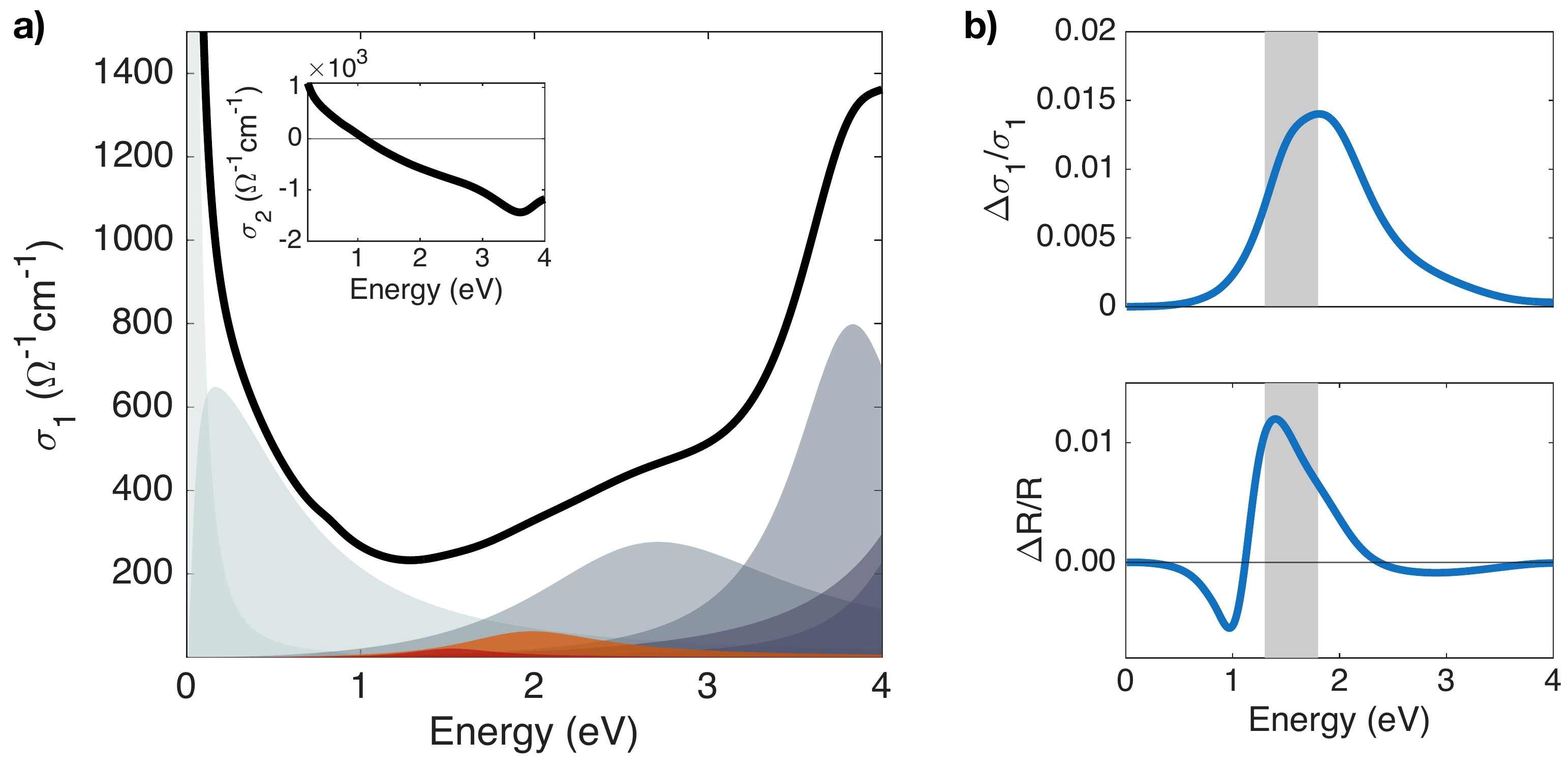}
\caption{(a) Optical conductivity of Y-Bi2212, parametrized using a sum of Drude-Lorentz oscillators, plotted as filled areas, following Ref. \citenum{Giannetti2011} and \citenum{Montanaro2024}. Main panel: real part $\sigma_1$. Inset: imaginary part $\sigma_2$. (b) Simulated variation of the optical response upon a $6\%$ increase in the spectral weight of the oscillators highlighted in panel (a) (red and orange areas). Top: relative change in absorption, $\Delta\sigma_1/\sigma_1$. Bottom: corresponding reflectivity variation, $\Delta R/R$. The gray shaded area marks the experimental spectral range, where an increase in absorption ($\Delta \sigma_1 > 0$) results in a positive transient reflectivity signal ($\Delta R/R > 0$).}
\label{fig:sign reflectivity}
\end{figure*}

Figure~\ref{fig:sign reflectivity}a reports the equilibrium optical conductivity of Y-Bi2212, adapted from Ref.~\citenum{Giannetti2011}, highlighting the characteristic spectral features of the cuprates. In particular, the onset of the charge-transfer excitations occurs at energies above $\sim 2.5$~eV, while the visible/near-infrared region probed in the present experiment (1.3–1.8~eV) lies below this threshold. In this spectral range, the optical response is dominated by weak transitions, whose microscopic origin is still under debate.

To assess whether the positive sign of the transient reflectivity signal ($\Delta R/R > 0$) observed in the 2DES experiment is compatible with a pump-induced increase in absorption, we model the equilibrium optical properties using a parametrization of the dielectric function based on a sum Drude–Lorentz, following Ref.~\citenum{Montanaro2024}. 
Within this framework, we simulate a photoinduced modification of the optical response by increasing the spectral weight of the oscillators closest to the probed energy window. Specifically, we consider a $6\%$ increase in the spectral weight of the oscillators highlighted in Fig.~\ref{fig:sign reflectivity}a, which predominantly contribute to the 1.3–1.8~eV range. From the modified dielectric function, we compute both the change in the real part of the optical conductivity, $\Delta \sigma_1$ (Fig. \ref{fig:sign reflectivity}b top panel), and the corresponding variation in reflectivity, $\Delta R/R$ (Fig. \ref{fig:sign reflectivity}b bottom panel).
Notably, an increase in absorption ($\Delta \sigma_1 > 0$) leads to a positive reflectivity variation ($\Delta R/R > 0$) within the spectral range probed in the experiment (highlighted by the gray area). This demonstrates that the experimentally observed positive transient reflectivity signal is fully consistent with a pump-induced enhancement of absorption due to the excitation of paramagnons.

\newpage
\section{Line profiles of the 2D spectra}

\begin{figure*}[h]
\includegraphics[width=12cm]{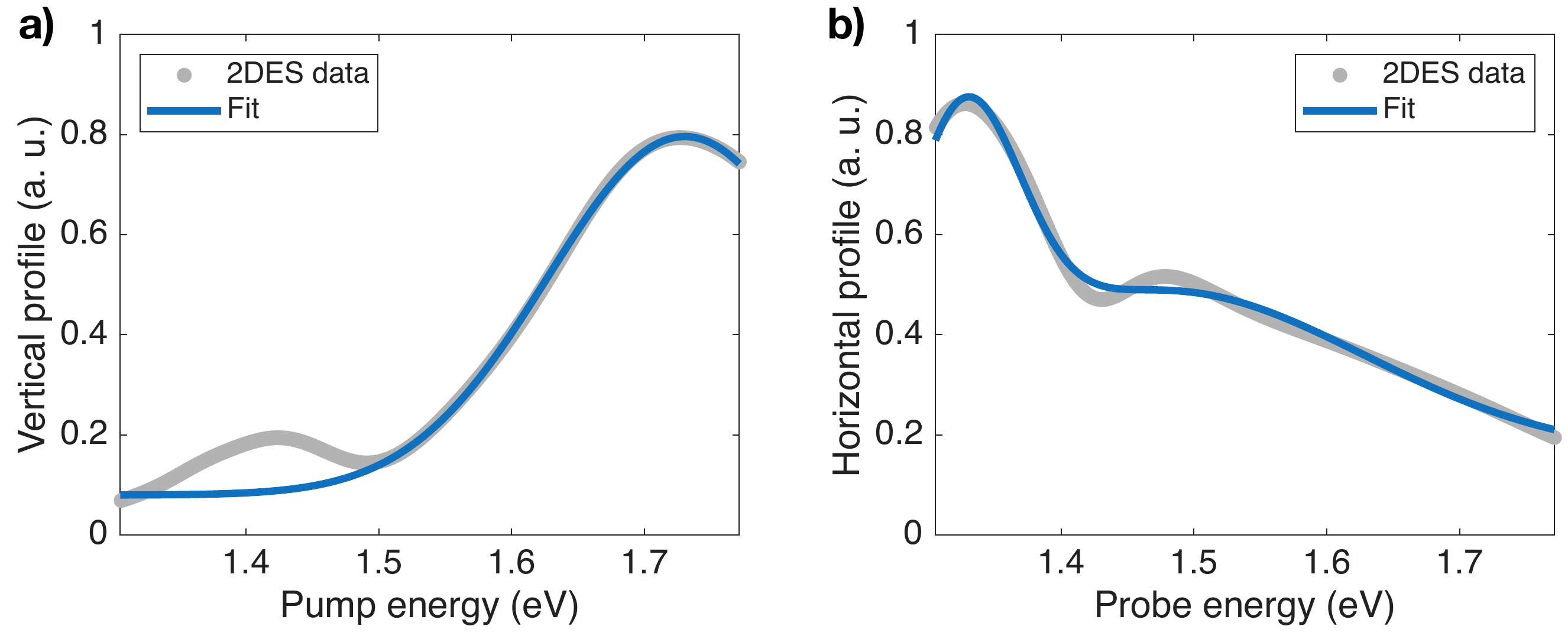}
\caption{Fitting of vertical (a) and horizontal (b) profiles of the 2DES map shown in Fig.~2a. The profiles are obtained by integrating the signal along $\hbar\omega_{\mathrm{pr}}$ between 1.3 and 1.4~eV, and along $\hbar\omega_{\mathrm{pu}}$ between 1.67 and 1.77~eV, respectively. The fits are performed using one or two Gaussian peaks to estimate the position of the off-diagonal peak ($(\hbar\omega_{\mathrm{pr}}, \hbar\omega_{\mathrm{pu}}) = (1.32, 1.73)$~eV).}
\label{fig:OP profiles fit}
\end{figure*}

To determine the position of the off-diagonal feature in the $(\hbar\omega_{\mathrm{pr}}, \hbar\omega_{\mathrm{pu}})$ plane, we extract horizontal and vertical line profiles from the 2DES map. Specifically, the horizontal line profile is obtained by integrating the signal over the pump-energy range $\hbar\omega_{\mathrm{pu}} = 1.67$--$1.77$~eV, while the vertical line profile is obtained by integrating over the probe-energy range $\hbar\omega_{\mathrm{pr}} = 1.3$--$1.4$~eV.
The resulting line profiles for the 2DES map shown in Fig.~2a are reported in Fig.~\ref{fig:OP profiles fit}. To extract the peak position, the horizontal line profile is fitted with the sum of two Gaussian functions: one accounting for the off-diagonal feature and a second describing the weak and broad residual signal extending along the diagonal in the $1.4$--$1.8$~eV energy range. From this fit, we find that the off-diagonal peak is centered at probe-energy $\hbar\omega_{\mathrm{pr}} = 1.32$~eV, and has a full width at half maximum of 100~meV.
The vertical line profile is instead fitted with a single Gaussian function, yielding the pump-energy position of the off-diagonal feature at $\hbar\omega_{\mathrm{pu}} = 1.73$~eV and a corresponding width of 240~meV.


\section{2DES fluence dependence}

\begin{figure*}[h]
\includegraphics[width=18cm]{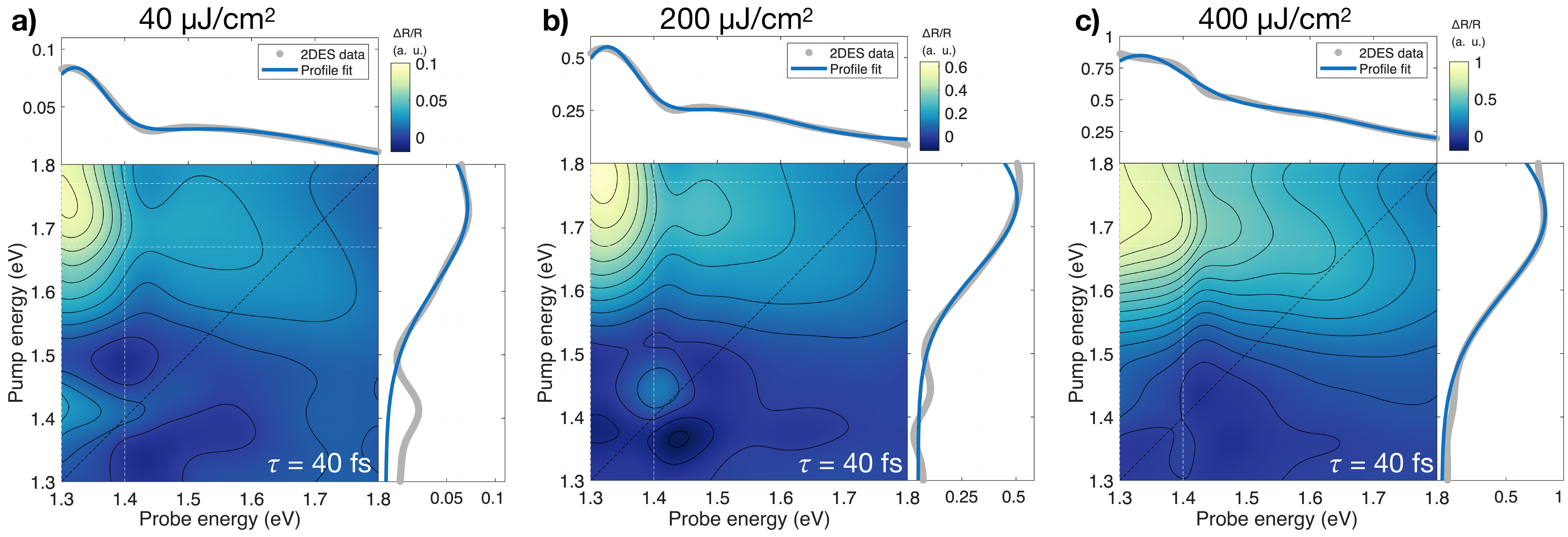}
\caption{2DES maps measured at room temperature on the optimally doped sample using excitation fluences of (a) 40 \textmu\text{J/cm}$^2$, (b) 200 \textmu\text{J/cm}$^2$, and (c) 400 \textmu\text{J/cm}$^2$.}
\label{fig:fluence maps}
\end{figure*}

Figure~\ref{fig:fluence maps} shows the fluence dependence of the 2DES maps measured on the optimally doped sample ($T_c = 95$~K) at room temperature and $\tau = 40$~fs. The excitation fluence is varied from $40$ \textmu\text{J/cm}$^2$ to $400$ \textmu\text{J/cm}$^2$. The off-diagonal feature is visible at all fluences, with its spectral position remaining unchanged. The signal amplitude scales with fluence, consistent with a linear increase of the photoexcited carrier density in this regime.

\newpage
\section{Coherent artifact}
\begin{figure}[h]
    \centering
    \includegraphics[width=0.35\textwidth,keepaspectratio]{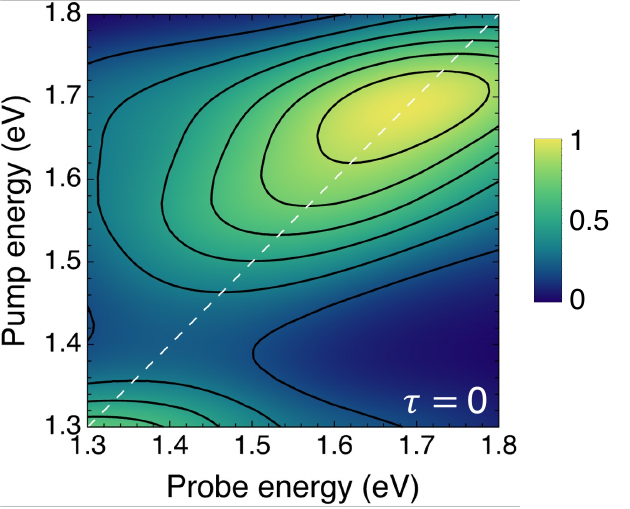}
    \caption{Calculated two-dimensional correlation map given by the pulse overlap at $\tau=0$.}
    \label{overlap1}
\end{figure}

To rule out the possibility that the observed 2DES signal originates from a coherent artifact due to the temporal overlap of pump and probe, we simulate the nonlinear response expected from pulse overlap alone.
The third-order nonlinear response mediated by a susceptibility $\chi^{(3)}(\omega_1,\omega_2,\omega_3)$ can be generically written as the convolution of three electric fields $\text{E}(\omega)$,
\begin{align}
    S(\hbar\omega_{pr})\propto\int d\omega_1d\omega_2d\omega_3 \,\chi^{(3)}(\omega_1,\omega_2,\omega_3)\text{E}(\omega_1)\text{E}(\omega_2)\text{E}(\omega_3)\,\delta(\omega_{pr}-\omega_1-\omega_2-\omega_3).
\end{align}
The dependence on the delay $t_{pu}$ between the two pump fields, $\text{E}_{pu1}$ and $\text{E}_{pu2}$, and on the delay $\tau$ between the second pump and the probe field $\text{E}_{pr}$ can be introduced by writing the total electric field as
\begin{align}
    \text{E}(\omega)\to\text{E}(\omega,t_{pu},\tau)=\text{E}_{pu1}(\omega)e^{-i\omega(t_{pu}+\tau)}+\text{E}_{pu2}(\omega)e^{-i\omega\tau}+\text{E}_{pr}(\omega).
\end{align}
With this substitution, the signal becomes $S(\hbar\omega_{pr},t_{pu};\tau)$, which reads
\begin{align}
    S(\hbar\omega_{pr},t_{pu};\tau)\propto6\int d\omega_1d\omega_2d\omega_3\,\chi^{(3)}(\omega_1,\omega_2,\omega_3)\text{E}_{pu1}(\omega_1)\text{E}_{pu2}(\omega_2)\text{E}_{pr}(\omega_3)e^{-i(\omega_1+\omega_2)\tau}e^{-i\omega_1t_{pu}}\,\delta(\omega_{pr}-\omega_1-\omega_2-\omega_3).
\end{align}
The 2DES signal is obtained by taking the Fourier Transform with respect to $t_{pu}$ \cite{Fiore2025}, 
\begin{align}\label{2dsig}
    S(\hbar\omega_{pr},\hbar\omega_{pu};\tau)&\propto6\int dt_{pu} e^{i\omega_{pu}t_{pu}}S(\hbar\omega_{pr},t_{pu};\tau)\nonumber\\
    &=6\int d\omega^\prime\, \chi^{(3)}(\omega_{pu},\omega^\prime,\omega_{pr}-\omega_{pu}-\omega^\prime)\text{E}_{pu1}(\omega_{pu})\text{E}_{pu2}(\omega^\prime)\text{E}_{pr}(\omega_{pr}-\omega_{pu}-\omega^\prime)e^{-i(\omega_{pu}+\omega^\prime )\tau}.
\end{align}
To model the coherent artifact, we consider only off-resonant electronic transitions, such that the third-order susceptibility is real and essentially frequency-independent, $\chi^{(3)}(\omega_{pu},\omega^\prime,\omega_{pr}-\omega_{pu}-\omega^{\prime})\sim 1$. In this case, Eq.\ \eqref{2dsig} reads
\begin{align}
    S(\hbar\omega_{pr},\hbar\omega_{pu};\tau)\propto6\,\text{E}_{pu1}(\omega_{pu})e^{-i\omega_{pu}\tau}\int d\omega^\prime\, \text{E}_{pu2}(\omega^\prime)\text{E}_{pr}(\omega_{pr}-\omega_{pu}-\omega^\prime)e^{-i\omega^\prime\tau}.
\end{align}
Fig.\ \ref{overlap1} shows the resulting 2DES signal for $\tau=0$, computed using the experimental pulse spectrum shown in Fig.~\ref{fig:setup}. The dominant feature is a broad peak along the diagonal $\hbar\omega_{pu}=\hbar\omega_{pr}$, with no narrow response peaked at $(1.3,1.7)$~eV. This demonstrates that the off-diagonal resonance observed experimentally cannot be attributed to a trivial coherent artifact originating from pump and probe overlap. We note that the amplitude of this artifact is governed by the temporal overlap between the pump and probe pulses and therefore vanishes on a timescale comparable to the pulse duration, whereas the off-diagonal feature in the experimental data persists well beyond this window, as shown in Fig.~2b of the main text.

%
%
%

\newpage
\section{Model of paramagnon-assisted indirect optical transitions}
\subsection{General framework}
To interpret the off-diagonal resonance observed in the 2DES spectra, we develop a model for indirect optical transitions in which an electron is promoted from an oxygen-derived valence state to the Cu-$3d_{x^2-y^2}$ conduction band with the assistance of a paramagnon that carries the momentum and energy required by conservation laws. The calculation is formulated in terms of three ingredients: (i) an initial-state spectral function $A_i(E,\mathbf{k}_i)$ for the O-$2p_{\pi,\sigma}$ bands involved in the charge-transfer excitation, where $\mathbf{k}_i$ is the initial momentum; (ii) a final-state spectral function $A_f(E,\mathbf{k}_f)$ for the Cu-$3d_{x^2-y^2}$ band, where $\mathbf{k}_f$ is the final momentum;  (iii) the imaginary part of the spin susceptibility $\chi''_{\mathbf q}(\hbar\Omega)$, which encodes the energy-momentum phase space available to the paramagnon mediating the transition, where $\mathbf{q}$ is the paramagnon momentum. For each photon energy value $\hbar\omega$, we compute a momentum-resolved transition probability $P_{if}(\hbar\omega,\mathbf{k}_f)$ and obtain the total probabilities for the paramagnon-emission (pump) and paramagnon-absorption (probe) channels by integration over the Brillouin zone. The resulting energy dependences are then combined to construct a simulated 2DES correlation map, which is directly compared to the experimental data.

We consider an indirect optical transition from an initial electronic state $\ket{i}$ to a final electronic state $\ket{f}$ driven by a photon of energy $\hbar\omega$ and assisted by the emission of a paramagnon carrying energy $\hbar\Omega_{\mathbf q}$ and momentum $\mathbf q$. Momentum conservation enforces
\begin{equation}
\mathbf{k}_f = \mathbf{k}_i + \mathbf{q}.
\label{eq:S3_momcons}
\end{equation}
Within the Fermi's Golden rule formalism, the transition probability (up to an overall prefactor) contains the energy-conservation constraint
\begin{equation}
P_{if}\propto\delta\!\left(E_f(\mathbf{k}_f)-E_i(\mathbf{k}_i)-\hbar\omega+\hbar\Omega_{\mathbf q}\right),
\label{eq:S3_onedelta}
\end{equation}
To separate the electronic and bosonic contributions, we introduce two auxiliary energy variables $E$ and $E'$ and use the identity
\begin{align}
\delta\!\left(E_f(\mathbf{k}_f)-E_i(\mathbf{k}_i)-\hbar\omega+\hbar\Omega_{\mathbf q}\right) = \int dE \int dE'\;
\delta\!\left(E-E_i(\mathbf{k}_i)\right)\;
\delta\!\left(E'-E_f(\mathbf{k}_f)+\hbar\omega\right)\;
\delta\!\left(E'-E+\hbar\Omega_{\mathbf q}\right).
\label{eq:S3_threedeltas}
\end{align}
Equation~\eqref{eq:S3_threedeltas} corresponds to the limit of vanishing lifetime broadening, in which both electronic states and bosonic excitations are assumed to be localized in energy. However, finite lifetimes and band dispersion broaden these constraints \cite{Grosso2013}. This is accounted for by replacing the electronic $\delta$-functions with the corresponding spectral functions, defined as the imaginary part of the retarded single-particle Green’s functions,
\begin{equation}
A_{i,f}(E,\mathbf{k}) = -\frac{1}{\pi}\,\mathrm{Im}\,G^{R}_{i,f}(E,\mathbf{k}).
\end{equation}
Analogously, the bosonic $\delta$-function enforcing the energy transfer $\hbar\Omega_{\mathbf q}$ is replaced by the bosonic spectral density, which for magnetic excitations is proportional to the imaginary part of the spin susceptibility, $\chi''_{\mathbf q}(\hbar\Omega)$. This procedure provides a generalization of Eq.~\eqref{eq:S3_threedeltas} that incorporates lifetime effects while preserving energy and momentum conservation on average. Within this spectral representation, the transition probability for an indirect, boson-assisted optical excitation driven by a photon of energy $\hbar\omega$ can be written as
\begin{align}
P_{if}(\hbar\omega,\mathbf{k}_f)\propto
\int\!\!\int dE\,dE'\;&
f(E_i)\,
\bigl[1-f(E_f)\bigr]\,
A_i\!\left(E - E_i(\mathbf{k}_i)\right)\,
A_f\!\left(E' - E_f(\mathbf{k}_f) + \hbar\omega\right)\,
\chi''\!\left(E' - E + \hbar\Omega_{\mathbf q}\right).
\label{eq:S3_finalP}
\end{align}
 The Fermi–Dirac distribution functions account for the electronic occupation of the initial and final states, respectively. The factor $f(E_i)$ enforces that the initial electronic state is occupied prior to photoexcitation, while the factor $[1-f(E_f)]$ ensures the availability of the final electronic state after the combined absorption of the photon and exchange of energy with the bosonic excitation. Here $E_f$ is already shifted by the Fermi energy.
\newline
\bigskip
\subsubsection{\textbf{Initial electronic state}}
The initial electronic state $\ket{i}$ represents an oxygen-derived valence state involved in the charge-transfer excitation. O-$2p_\pi$ and O-$2p_\sigma$ bands are described by a Lorentzian spectral function,
\begin{equation}
A_i(E,\mathbf{k}) =
\frac{\eta}{\left(E - E_i(\mathbf{k})\right)^2 + \eta^2},
\label{eq:S3_Ai}
\end{equation}
where the broadening is set to $\eta = k_BT=24$ meV corresponding to thermal broadening at room temperature ($T=300$~K), with $k_B$ the Boltzmann constant. The dispersions of the two oxygen bands are parametrized as
\begin{equation}
E_{2p_\pi}(\mathbf{k}) = -0.30\!\left[(k_x-\pi)^2+(k_y-\pi)^2\right]-1.3~\mathrm{eV},
\qquad
E_{2p_\sigma}(\mathbf{k}) = -0.30\!\left(k_x^2+k_y^2\right)-1.3~\mathrm{eV}.
\label{eq:S3_Ei}
\end{equation}
The curvature parameter ($-0.30$) and the energy offset ($-1.3$~eV) are obtained by fitting the oxygen-derived valence bands reported in Ref.~\onlinecite{Cilento2018}.
In the calculations, the initial momentum is fixed at $\mathbf{k}_i=(\pi,\pi)$ for the O-$2p_\pi$ band and at $\mathbf{k}_i=(0,0)$ for the O-$2p_\sigma$ band, corresponding to the maxima of their spectral weight.
\newline
\bigskip
\subsubsection{\textbf{Final electronic state}}
The final electronic state $\ket{f}$ corresponds to the Cu-$3d_{x^2-y^2}$ conduction band. Its spectral function is obtained from the single-particle Green’s function of an extended Hatsugai--Kohmoto model, which incorporates short-range antiferromagnetic correlations through a momentum-space coupling between $\mathbf{k}$ and $\mathbf{k}+\mathbf{\Pi}$, with $\mathbf{\Pi}=(\pi,\pi)$ \cite{Worm2024}. In this framework, the electronic occupation factor $n_{\mathbf{k}+\mathbf{\Pi}}$ encodes the redistribution of spectral weight induced by the $\mathbf{\Pi}$-coupling and is computed self-consistently following the procedure described in Ref.~\cite{Worm2024}. The retarded Green’s function reads
\begin{equation}
G_{\mathbf{k}}(E)=
\frac{1-n_{\mathbf{k}+\mathbf{\Pi}}}{E+\mu-\epsilon(\mathbf{k})+i\eta}
+
\frac{n_{\mathbf{k}+\mathbf{\Pi}}}{E+\mu-\epsilon(\mathbf{k})-\mathcal{V}+i\eta},
\label{eq:S3_Gf}
\end{equation}
where $\mathcal{V}$ is the interaction parameter controlling the splitting between the two branches, $\mu$ is the chemical potential, and $\eta$ is the same as for the initial state. The final energy entering Eq. \ref{eq:S3_finalP} is $E_f=\epsilon(\mathbf{k})-\mu$. The corresponding spectral function is obtained as
\begin{equation}
A_f(E,\mathbf{k}) = -\frac{1}{\pi}\,\mathrm{Im}\,G_{\mathbf{k}}(E).
\end{equation}
The bare dispersion $\epsilon(\mathbf{k})$ entering Eq.~\eqref{eq:S3_Gf} is taken as a tight-binding form,
\begin{equation}
\epsilon(\mathbf{k}) =
-2t(\cos k_x + \cos k_y)
-4t'\cos k_x \cos k_y
-2t''(\cos 2k_x + \cos 2k_y),
\label{disp_tb}
\end{equation}
with parameters $t=0.4$~eV, $t'=-0.08$~eV, and $t''=0.04$~eV \cite{Norman2006, Damascelli2003, Comin2014}. 
In order to anchor the theoretical Fermi surface entering the final-state Green’s function in Eq.~\eqref{eq:S3_Gf}, we need to fix the chemical potential $\mu$ and the interaction parameter $\mathcal{V}$. To estimate these quantities, we performed ARPES measurements of the Fermi surface of optimally doped Y-Bi2212. These experiments were performed at the Quantum Materials Spectroscopy Centre (QMSC) beamline of the Canadian Light Source (CLS), using a Scienta R4000 hemispherical analyzer with angular and energy resolutions better than 0.1$^\circ$ and 15 meV respectively. The ARPES data were acquired using linear vertical polarized light at a photon energy of 88 eV, at a base pressure lower than $5\times10^{-11}$ Torr and a temperature of 14 K. This is well below the $T_c$ of these samples and therefore we probe the electronic structure in the superconducting state. As a consequence, spectral weight at the antinodes is strongly suppressed by the superconducting gap, and only the ungapped nodal arcs are directly visible at $E_{Fermi}$. The selected samples were aligned along the nodal $\Gamma$-Y direction where a structural supermodulation in the Bi–O planes gives rise to multiple diffraction replicas of the main bands in momentum space. These replicas are clearly visible in the experimental data but do not modify the underlying Fermi-surface topology. For the purpose of the present work, the ARPES measurement is used exclusively to fix the parameters entering the final-state electronic structure. In particular, the chemical potential $\mu$ and the interaction parameter $\mathcal{V}$ are determined by matching the nodal Fermi momenta of the model dispersion to those extracted from ARPES. This procedure yields $\mu = -0.448$~eV and $\mathcal{V} = 0.076$~eV.
\begin{figure}
    \centering
    \includegraphics[width=0.8\linewidth]{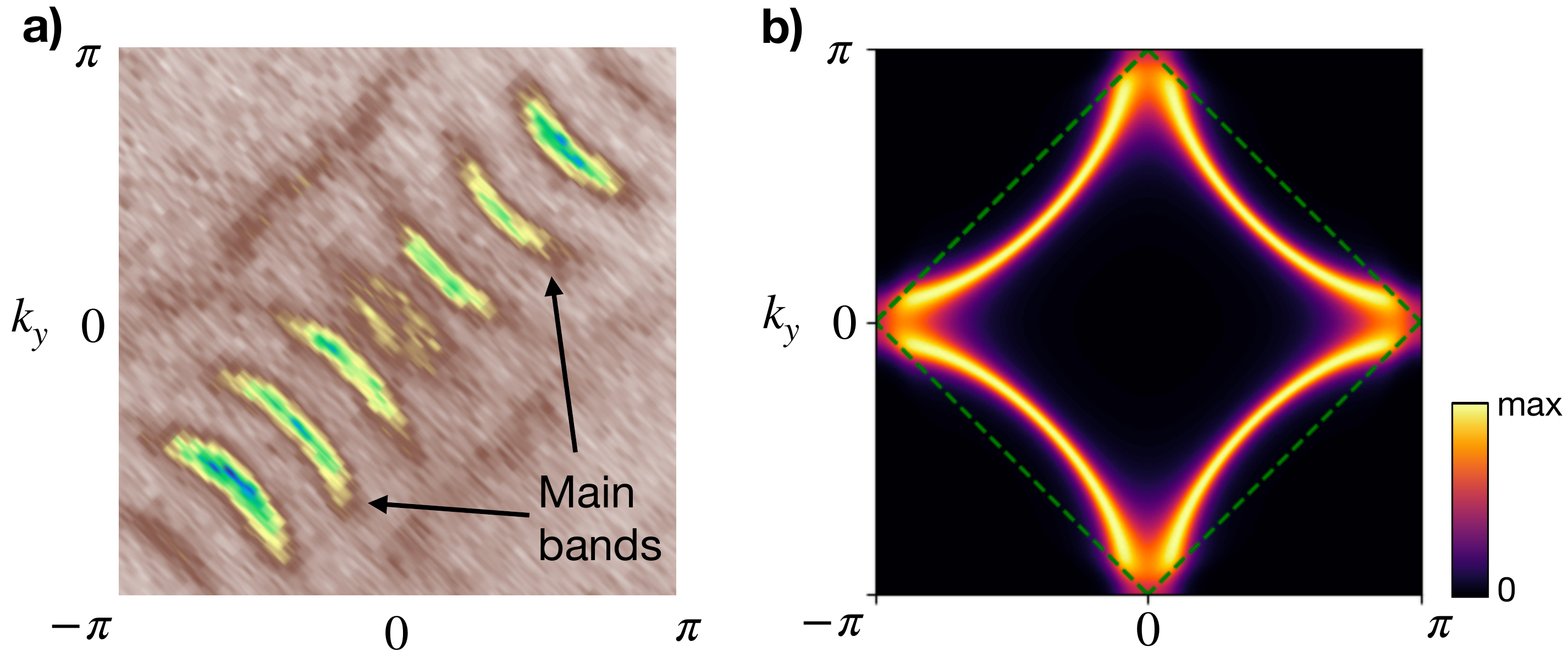}
 \caption{(a) Fermi-surface map of optimally doped Bi$_2$Sr$_2$Ca$_{0.92}$Y$_{0.08}$Cu$_2$O$_{8+\delta}$ measured by ARPES at $T \simeq 14$~K. The main Fermi-surface bands are indicated, while additional features arise from diffraction replicas due to the Bi--O superstructure. 
(b) Calculated Fermi surface obtained from the final-state electronic structure used in the model with $\mu = -0.448$~eV and $\mathcal{V} = 0.076$~eV . Green dashed lines indicate the magnetic Brillouin zone.}
    \label{fig:placeholder}
\end{figure}
\newline
\bigskip
\subsubsection{\textbf{Magnetic susceptibility}}
The bosonic excitation assisting the indirect optical transition is modeled as a paramagnon, described by the imaginary part of the spin susceptibility $\chi''_{\mathbf q}(\hbar\Omega)$. We adopt a phenomenological form commonly used to describe damped antiferromagnetic spin fluctuations in cuprates, consistent with resonant inelastic x-ray scattering (RIXS) measurements \cite{LeTacon2011, Peng2018}. The paramagnon dispersion is taken as
\begin{equation}
\Omega_{\mathbf q} =
\frac{J}{\hbar} \sqrt{\left(2-\cos q_x-\cos q_y\right)
\left(2+\cos q_x+\cos q_y+\frac{\omega_{\mathrm{AF}}^2}{4J^2}\right)},
\end{equation}
where $J$ is the antiferromagnetic exchange interaction and $\omega_{\mathrm{AF}}$ sets a possible spin-gap at the antiferromagnetic wave vector. In the present calculations we set $\omega_{\mathrm{AF}}=0$, appropriate for short-range antiferromagnetic correlations in the doped compound. The corresponding susceptibility is modeled as
\begin{equation}
\chi''_{\mathbf q}(\hbar\Omega) =
\frac{2\Gamma\,\hbar\Omega\left[1-\tfrac{1}{2}(\cos q_x+\cos q_y)\right]}
{\left[(\hbar\Omega)^2-(\hbar\Omega_{\mathbf q})^2\right]^2+4\Gamma^2\hbar^2\Omega^2},
\label{eq:S3_chi}
\end{equation}
where $\Gamma$ is a phenomenological damping parameter accounting for the finite lifetime of the magnetic excitations. The parameters used in the calculations are $J=105$~meV \cite{LeTacon2011,Dean2013,Dean2013_PRL,Dahm2009} and $\Gamma=300$~meV, in agreement with values reported by RIXS experiments on Bi-based cuprates \cite{Peng2018}. This form of $\chi''_{\mathbf q}(\hbar\Omega)$ captures both the dispersion and the strong damping characteristic of high-energy paramagnons in the normal state and provides the bosonic spectral density entering the transition probability in Eq.~\ref{eq:S3_finalP}.
\newline
\bigskip
\subsection{Separation of paramagnon emission and absorption channels}
In the present framework, emission and absorption processes correspond to distinct regions of the bosonic spectral function. The energy transferred to the magnetic subsystem is given by $\Omega = E' - E$, where $E$ and $E'$ are the electronic energies appearing in the convolution of spectral functions. Processes with $\Omega < 0$ correspond to the emission of a paramagnon during the optical transition, while $\Omega > 0$ describes paramagnon absorption.
\newline
In the numerical evaluation of the transition probability, these two contributions are treated separately by restricting the support of the susceptibility $\chi''_{\mathbf q}(\hbar\Omega)$ to negative or positive $\Omega$, depending on whether the pump (emission) or probe (absorption) process is considered. This procedure allows us to explicitly resolve the two branches of the indirect transition and to separately identify the momentum and energy phase space associated with paramagnon creation and annihilation.
\newline
\bigskip
\subsection{Numerical implementation}
The transition probability in Eq.~\eqref{eq:S3_finalP} was evaluated numerically by explicit integration over electronic energies and final-state momenta. The final electronic momentum $\mathbf{k}_f$ was sampled on a uniform $100\times100$ grid in the reduced Brillouin zone $[0,\pi]\times[0,\pi]$. For each $\mathbf{k}_f$, the transferred bosonic momentum is uniquely fixed by momentum conservation, $\mathbf{q}=\mathbf{k}_f-\mathbf{k}_i$.

Energy integrals were performed over a finite window $[-5,5]$~eV, well exceeding all electronic and bosonic energy scales relevant to the problem. Numerical integrations were carried out using adaptive routines with relative and absolute tolerances set to $10^{-8}$. The finite lifetime of electronic and bosonic excitations ensures convergence of the integrals through the broadening parameters entering the spectral functions.
\subsection{Results for the O-$2p_{\sigma}$ channel ($\mathbf{k}_i=(0,0)$)}
\label{sec:S3_results_k0}
In Fig.~4 in the main text we present the momentum-resolved probability maps for the indirect paramagnon-assisted transition evaluated for initial states at $\mathbf{k}_i=(\pi,\pi)$. For completeness, here we report the corresponding results obtained by fixing the initial momentum at $\mathbf{k}_i=(0,0)$, which selects the O-$2p_{\sigma}$ initial state (see Eq.~\eqref{eq:S3_Ei}). The probability $P_{if}(\hbar\omega,\mathbf{k}_f)$ is computed from Eq.~\eqref{eq:S3_finalP} under the same assumptions and parameters used in the main text.

\begin{figure*}[h]
    \centering
    \includegraphics[width=0.7\linewidth]{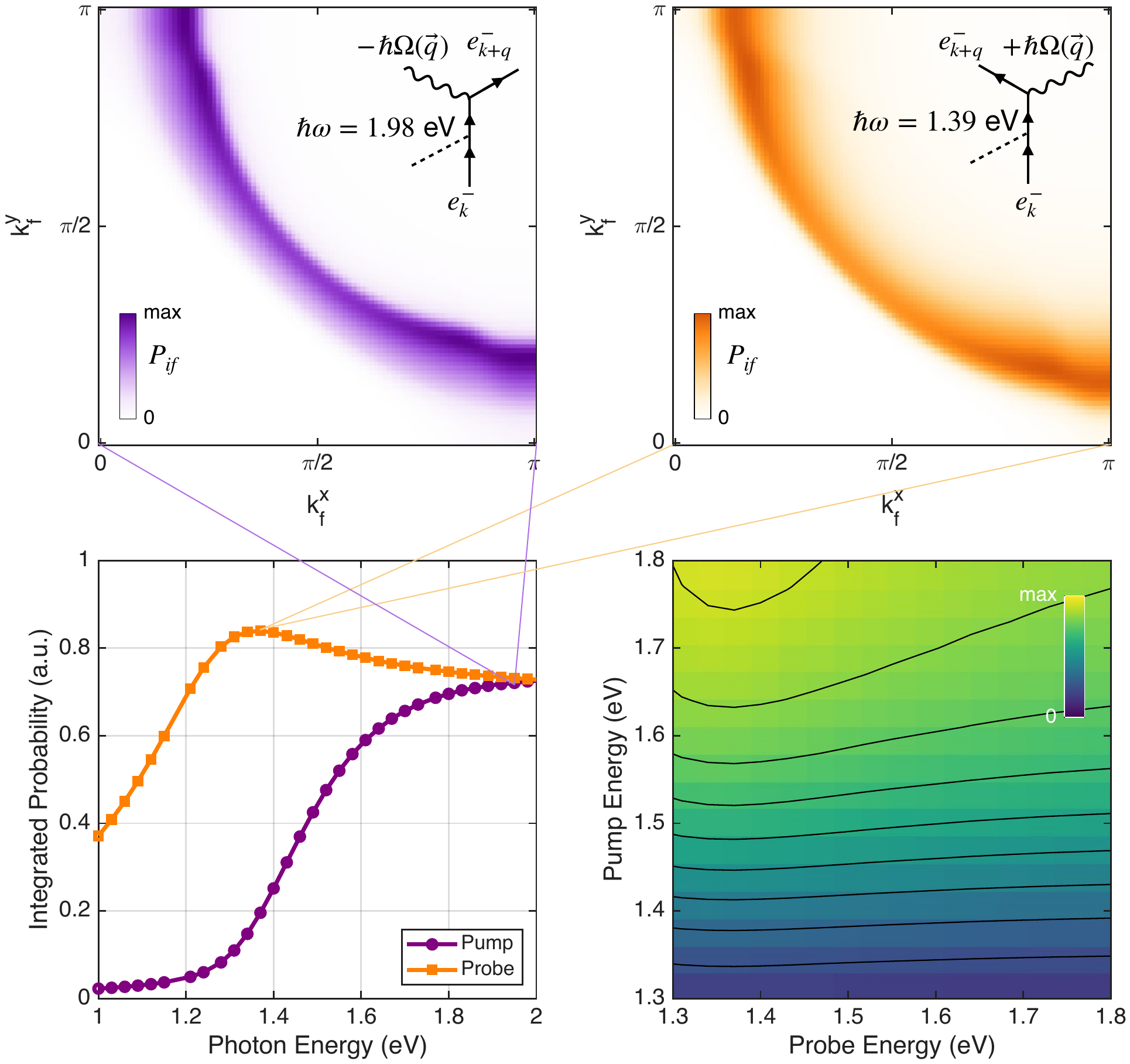}
    \caption{Momentum- and energy-conservation results for an initial state at $\mathbf{k}_i=(0,0)$ (O-$2p_{\sigma}$ channel).
    (a) Momentum-resolved probability $P_{if}(\mathbf{k}_f)$ for the pump process at fixed photon energy $\hbar\omega=1.98$~eV, corresponding to excitation followed by paramagnon emission, as sketched in the inset diagram.
    (b) Momentum-resolved probability $P_{if}(\mathbf{k}_f)$ for the probe process at fixed photon energy $\hbar\omega=1.39$~eV, involving absorption of a paramagnon previously generated by the pump (inset).
    (c) Momentum-integrated transition probabilities for pump (purple) and probe (orange) processes as functions of photon energy, obtained by integrating the maps in panels (a) and (b) over the Brillouin zone for each $\hbar\omega$.
    (d) Simulated two-dimensional correlation map constructed as the product of the pump and probe energy dependences shown in panel (c).}
    \label{fig:SI_results_k0}
\end{figure*}
\subsection{Paramagnon phase space for the probe (absorption) channel}
\label{sec:S3_qsum}
\begin{figure}
    \centering
    \includegraphics[width=0.5\linewidth]{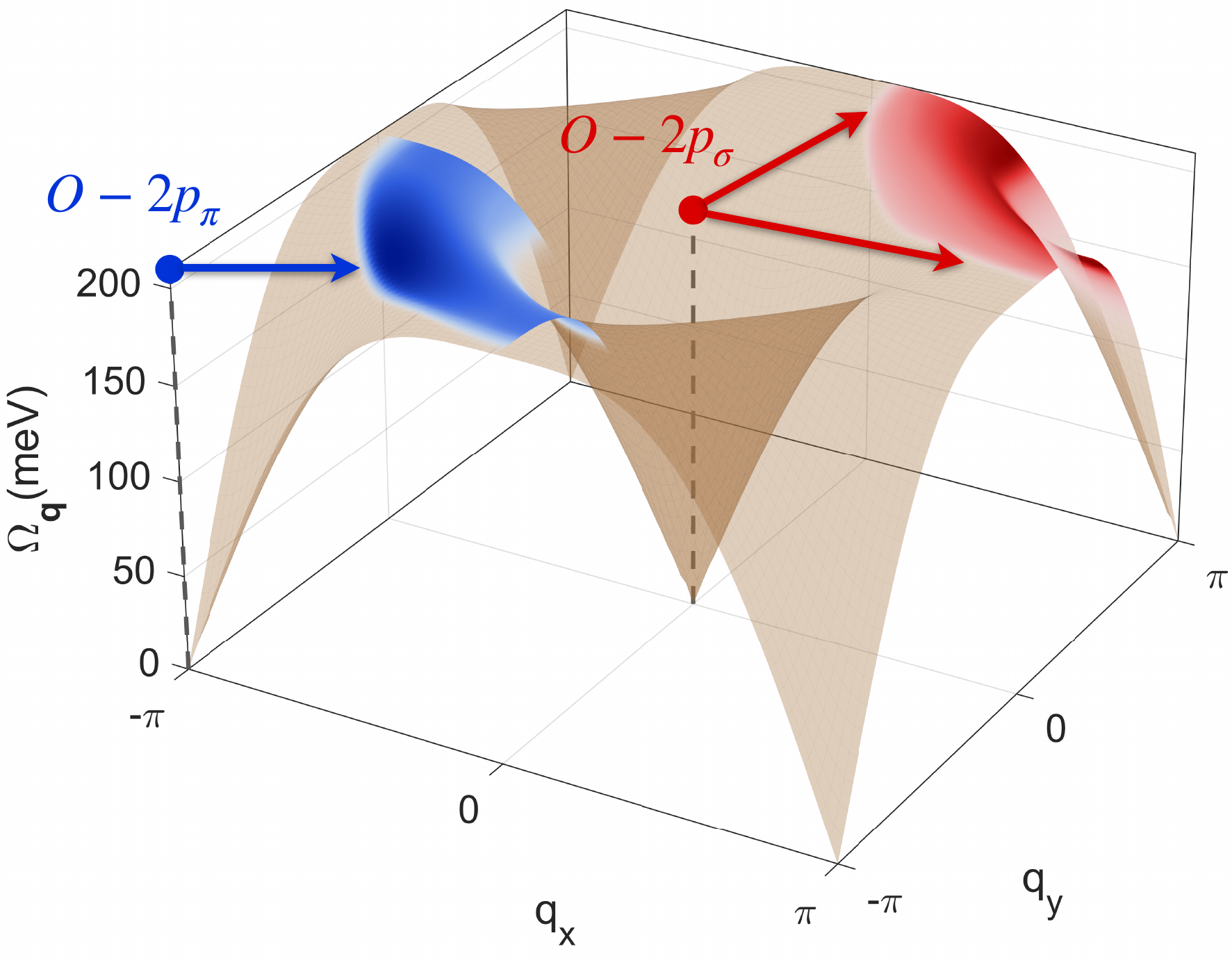}
    \caption{Antiferromagnetic (paramagnon) dispersion and momentum-transfer phase space for the absorption (probe) channel. The semi-transparent surface shows the model paramagnon dispersion $\Omega_{\mathbf{q}}$ over the full Brillouin zone. The blue [red] shaded areas highlight the set of momenta $\mathbf{q}$ contributing to the paramagnon absorption process for the initial O-$2p_{\pi}$ [O-$2p_{\sigma}$] state. For visualization, the $(\pi,\pi)$-centered contribution is translated by a reciprocal-lattice vector and displayed around $(-\pi,-\pi)$.}
    \label{fig:dispersion_abs}
\end{figure}

In the main text we show all the set of momenta involved in the paramagnon emission, here we report the analogous for the absorption process. In Fig.~\ref{fig:dispersion_abs} we plot the antiferromagnetic dispersion $\Omega_{\mathbf q}$ used in the calculations and highlight the set of momenta $\mathbf q$ that contribute to the absorption (probe) process. 
The transferred momentum is fixed by momentum conservation, $\mathbf q=\mathbf k_f-\mathbf k_i$, and the corresponding energy transfer is selected by the absorption channel. 
The blue [red] shaded region corresponds to the momenta exchanged when the initial O-$2p$ state is taken at $(\pi,\pi)$ [$(0,0)$]. 
For clarity, the $(\pi,\pi)$-centered contribution is translated by a reciprocal-lattice vector and displayed in the same zone as the $(0,0)$-centered one. 
We find that the absorption channel samples a different portion of the paramagnon phase space with respect to emission, while remaining concentrated near the upper part of the paramagnon band, consistent with the boson energy scale extracted from the off-diagonal 2DES correlation.

\clearpage
\section{2DES data for superconducting phase}
\begin{figure*}[h]
\includegraphics[width=16cm]{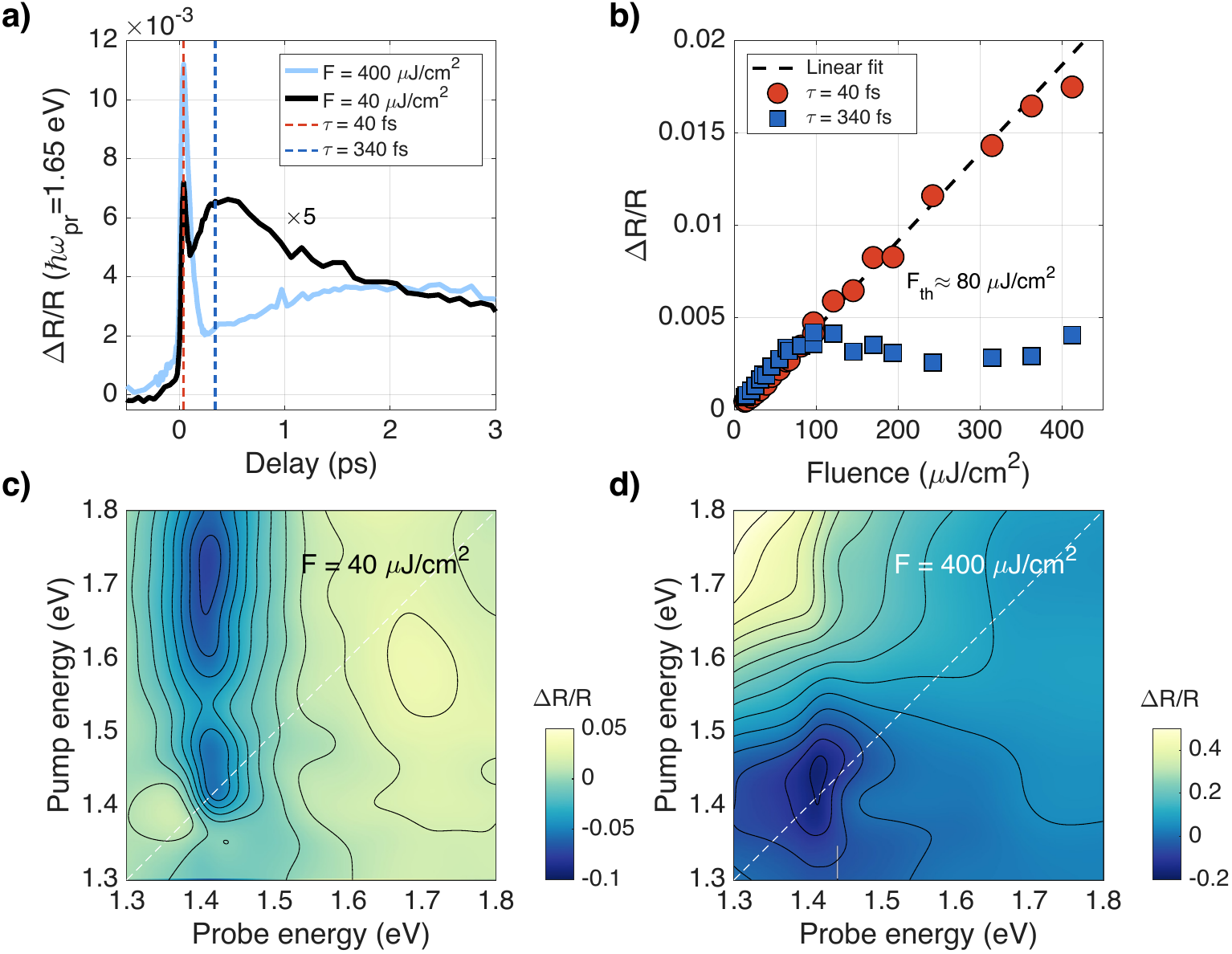}
\caption{{(a) Temporal dynamics of the differential reflectivity $\Delta R/R$ measured at $\hbar\omega_{\mathrm{pr}} = 1.65$~eV for two excitation fluences, $F = 40$~\textmu J/cm$^2$ ($\times 5$ magnification) and $F = 400$~\textmu J/cm$^2$. The dashed red and blue lines mark the delays $\tau = 40$~fs and $\tau = 340$~fs at which the fluence dependence in panel (b) is evaluated. (b) Fluence dependence of $\Delta R/R$ at $\hbar\omega_{\mathrm{pr}} = 1.65$~eV extracted at $\tau = 40$~fs (red circles) and $\tau = 340$~fs (blue squares). The dashed black line is a linear fit to the fast-component data. The threshold fluence $F_{\mathrm{th}} = 80$ \textmu\text{J/cm}$^2$ is indicated. (c, d) Two-dimensional $\Delta R/R$ maps at $\tau = 340$~fs as a function of pump and probe photon energy at low fluence ($F = 40$ \textmu\text{J/cm}$^2$) and high fluence ($F = 400$ \textmu\text{J/cm}$^2$), respectively. The white dashed line marks the diagonal ($\hbar\omega_{\mathrm{pu}} = \hbar\omega_{\mathrm{pr}}$).}}
\label{fig:temperature maps}
\end{figure*}

Figure~\ref{fig:temperature maps} reports measurements performed on the optimally 
doped sample below $T_c$. Panel (a) displays the temporal dynamics of $\Delta R/R$ at $\hbar\omega_{\mathrm{pr}} = 1.65$~eV, where a fast component peaking at 
$\tau = 40$~fs, and a slower component with a fluence-dependent relaxation 
dynamics can be identified. These two components display distinct fluence dependences, as shown in panel (b), which reports $\Delta R/R$ extracted at fixed delays $\tau = 40$~fs and $340$~fs. The fast component exhibits a linear fluence dependence, while the slower component reaches saturation at a threshold $F_{\mathrm{th}} \simeq 80$ \textmu\text{J/cm}$^2$, signaling the onset of photoinduced melting of the superconducting condensate, consistent with previous studies on the same compound \cite{Giannetti2009}. The 2DES response measured in the superconducting phase at fluence below $F_{\mathrm{th}}$, reported in Fig.~\ref{fig:temperature maps}c at $\tau = 340$~fs, exhibits markedly different spectral features compared to the normal-state 2DES spectrum (see Fig.~2b in the main text). Conversely, at high excitation fluence ($400$ \textmu\text{J/cm}$^2$), the normal-state response is recovered even at $T < T_c$ (Fig.~\ref{fig:temperature maps}d). A detailed analysis of the low-temperature regime is beyond the scope of the present work and will be addressed in a future study.

\clearpage
\newpage
\section{Coupling strength estimate}
\label{sec: coupling strength estimate}
The onset time of the off-diagonal resonance in 2DES spectra provides information about the coupling strength between the charge-transfer and spin excitations, with faster build-up times corresponding to stronger coupling. In our measurements, the off-diagonal feature appears essentially instantaneous within the duration of the pulses. Nevertheless, it is possible to estimate an upper bound on the build-up time of the off-diagonal resonance at $\hbar \omega_{pr} = 1.3$~eV. This, in turn, allows us to place a lower bound on the corresponding coupling strength. \\

\begin{figure*}[]
\includegraphics[width=15cm]{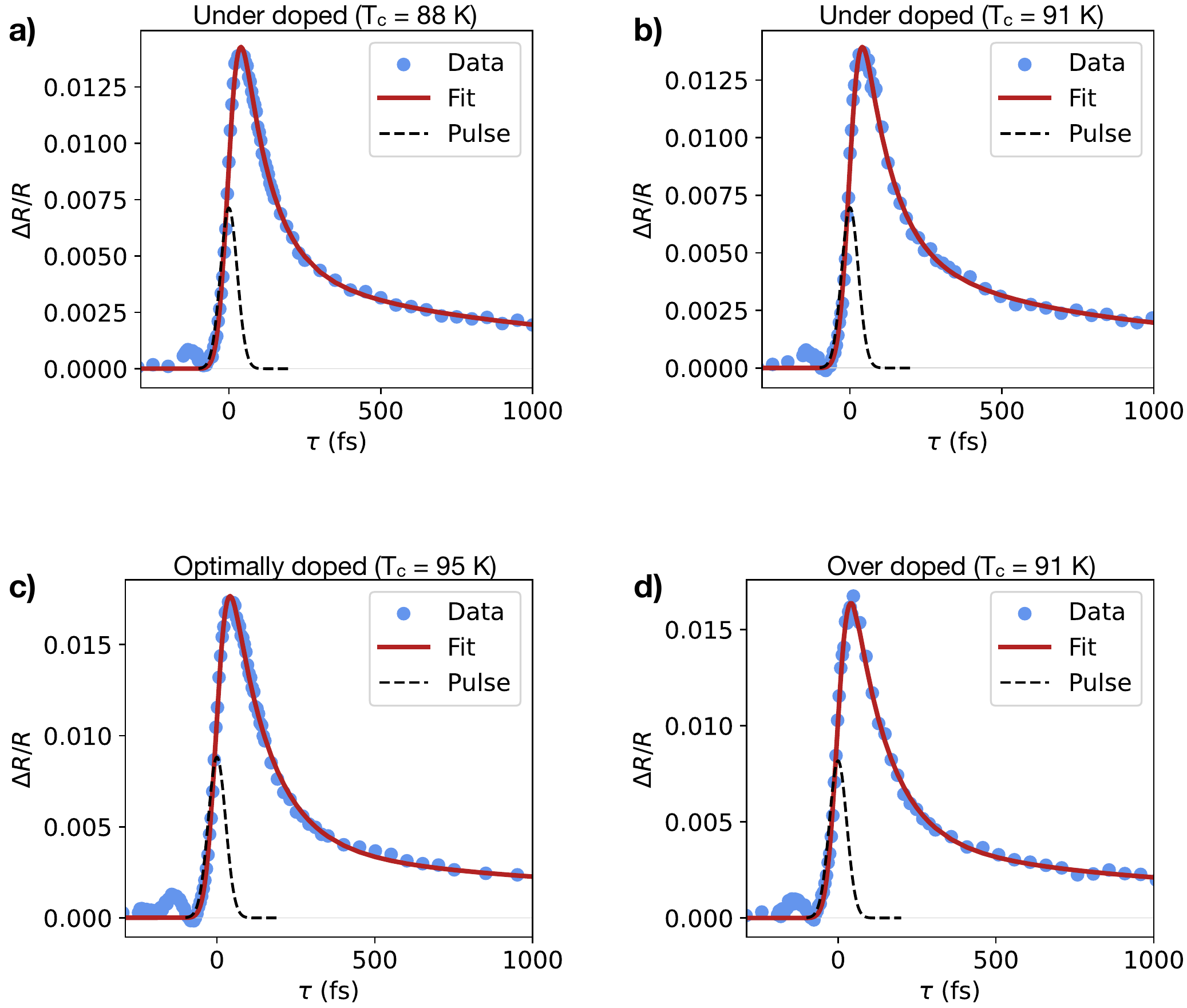}
\caption{Transient reflectivity dynamics measured at $\hbar \omega_{pr} = 1.3$~eV, representing the dynamics of the off-diagonal component in the 2D maps, along with corresponding double-exponential decay fits. The black dashed line shows the Gaussian accounting for the experimental pulse duration. Data were acquired at room temperature using a $400$ \textmu\text{J/cm}$^2$ excitation fluence on different samples: (a) underdoped ($T_c = 88$~K, $p=0.13$), (b) underdoped ($T_c = 91$~K, $p=0.14$), (c) optimally doped ($T_c = 95$~K, $p=0.16$), and (d) overdoped ($T_c = 91$~K, $p=0.18$).} 
\label{fig:fit dynamics}
\end{figure*}

The dynamics of the transient reflectivity response at $\hbar \omega_{pr} = 1.3$~eV are well described by a double-exponential decay function convoluted with the Gaussian of FWHM = 65 fs accounting for the pulses time duration. Fig. \ref{fig:fit dynamics} shows the corresponding fits for measurements performed on four samples with different doping levels. Introducing a third exponential to account for rise time does not result in a statistically significant improvement of the fit quality. Moreover, when the additional component is included, the extracted rise time and decay constants become unstable and poorly constrained, indicating that the data do not support the presence of a resolvable build-up within the temporal resolution of the experiment.

\begin{figure*}[]
\includegraphics[width=12cm]{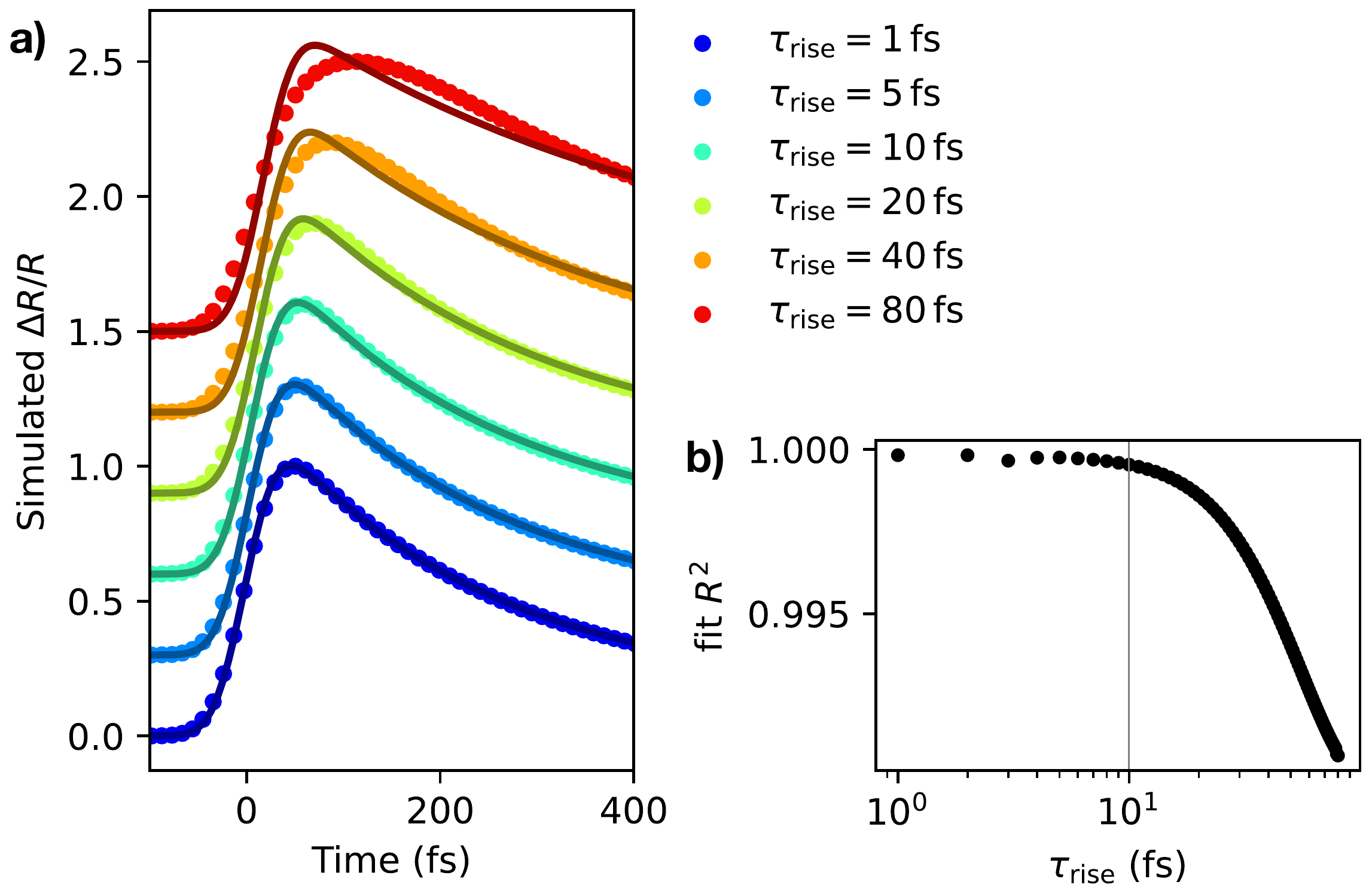}
\caption{(a) Synthetic dataset simulating transient reflectivity dynamics with a build-up time, $\tau_{\mathrm{rise}}$, followed by a double-exponential decay. Solid lines show fits using a double-exponential decay function, demonstrating that, given the experimental pulse duration, only build-up times longer than 10~fs would be detectable.
(b) Coefficient of determination ($R^2$) of the fits shown in panel (a), illustrating that the double-exponential fit accurately describes the dynamics for $\tau_{\mathrm{rise}} \lesssim 10$~fs.}
\label{fig:build up bound}
\end{figure*}

To quantify the shortest resolvable build-up time, we performed numerical simulations in which synthetic datasets were generated including an exponential rise component with varying time constant $\tau_{\mathrm{rise}}$, followed by two exponential decay components. The resulting signals were convoluted with the experimental pulse duration (Gaussian FWHM = $65$ fs). These synthetic datasets were then fitted using the same double-exponential decay model (without a rise component) employed for the analysis of the experimental data. As shown in Fig. \ref{fig:build up bound}a, for  $\tau_{\mathrm{rise}} \lesssim 10$ fs the fits accurately reproduce the simulated dynamics, and no systematic deviations are observed at early time delays. Only for longer rise times does the omission of a rise term lead to detectable discrepancies (see Fig. \ref{fig:build up bound}b).
We therefore conclude that any $\Delta R/R$ build-up occurs on a timescale faster than approximately 10 fs, which represents an upper bound set by the experimental time resolution. \\


\begin{figure*}[]
\includegraphics[width=16cm]{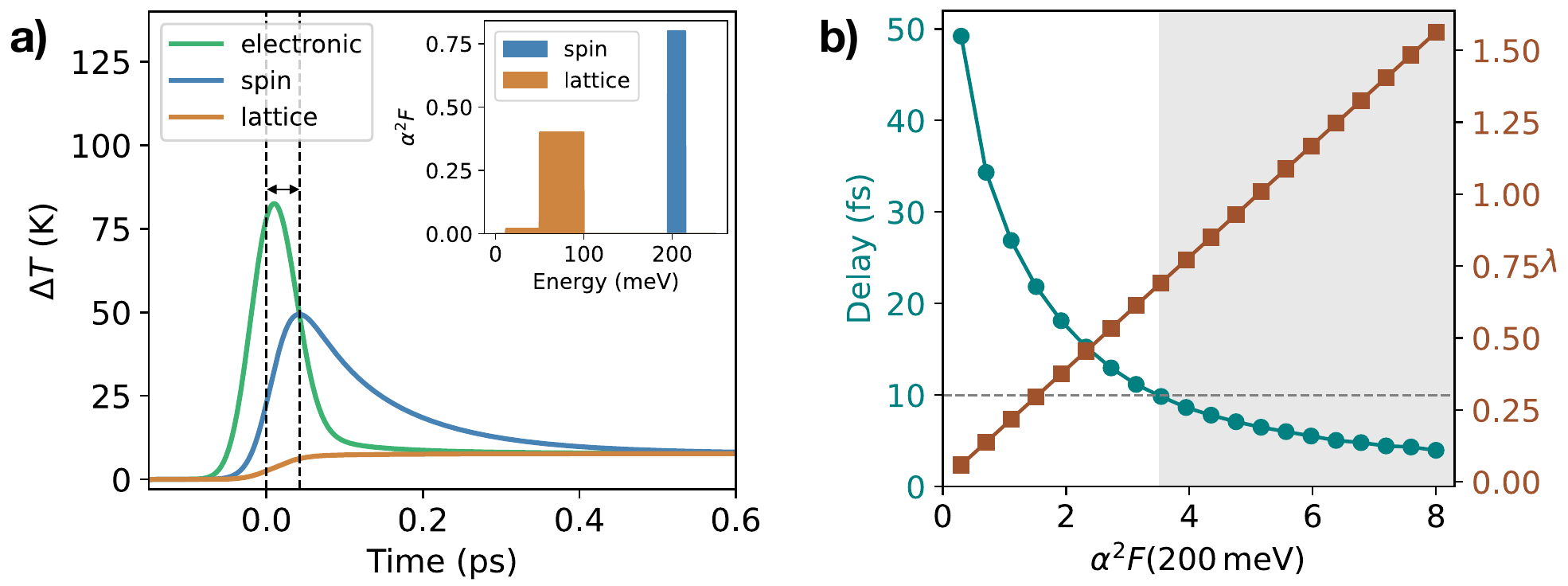}
\caption{(a) Time evolution of the electronic, spin, and lattice temperatures computed with the three-temperature model. Inset: spin and lattice contributions to the bosonic spectral function $\alpha^2F(\omega)$, represented as a sum of histograms.
(b) Value of the delay time at which the spin temperature reaches its maximum (left axis) as a function of the amplitude of $\alpha^2F(\omega)$ at 200~meV. The corresponding electron-spin coupling strength, $\lambda$, is shown on the right axis. The shaded gray area highlights the region where the spin response occurs faster than 10~fs (dashed line), which identifies the lower bound on the coupling ($\lambda \gtrsim 0.7$).}
\label{fig:3TM}
\end{figure*}

We now turn to estimating a lower bound for the coupling strength between charge excitations and paramagnons. To this end, we employ a three-temperature model describing the relaxation of the electronic (charge-transfer excitations) subsystem through its coupling to spin excitations at $200$ meV, and—on longer timescales—to the lattice. The coupling is encoded in a bosonic spectral function, $\alpha^2F(\omega)$, composed of a spin component peaked at 200~meV and a lattice (phononic) component spanning 10–100~meV (see inset of Fig.~\ref{fig:3TM}a).
Following photoexcitation, the rapid increase in electronic temperature drives a subsequent rise of the spin temperature and, at later times, of the lattice temperature - as modeled through coupled differential equations describing the energy transfer among the three population subsets \cite{DalConte2012,mor2025selective} - as illustrated in Fig.~\ref{fig:3TM}a. The dynamics depend on the specific heats of the three subsystems: the electronic specific heat is taken as $C_{\mathrm{el}}=\gamma T$ with $\gamma = 1\times 10^{-4}\,\mathrm{J\,cm^{-3}\,K^{-2}}$ and $T = 290$~K\cite{DalConte2012}; the lattice specific heat is $C_{lat} = 2.27\,\mathrm{J\,cm^{-3}\,K^{-1}}$~\cite{junod1994specific,DalConte2012}; and the spin specific heat is set to $C_{spin} = 0.029\,\mathrm{J\,cm^{-3}\,K^{-1}}$, consistent with the expectation that for bosonic modes of electronic-origin one has $C_{\mathrm{spin}}\lesssim 0.1 C_{\mathrm{el}}$~\cite{DalConte2012}. The pump excitation intensity is chosen to yield an electronic temperature increase of approximately $80$--$100\,\mathrm{K}$, consistent with the expected electronic heating considering the fluence used experimentally \cite{Boschini2018}.

Within this framework, we determine the spin coupling strength ($\alpha^2F(\omega=200\mathrm{~meV})$) that produces a delay of no more than $10\,\mathrm{fs}$ between the maxima of the electronic and spin temperatures, in accordance with the experimental upper limit. Figure~\ref{fig:3TM}b reports the dependence of this delay on the coupling strength, together with the corresponding electron-spin coupling strength parameter $\lambda = 2 \int_0^\infty \frac{\alpha^2F(\omega)}{\omega} d\omega $ as a function of $\alpha^{2}F(200\,\mathrm{meV})$. The experimentally constrained maximum delay of $10\,\mathrm{fs}$ (dashed line) sets a lower bound $\lambda \gtrsim 0.7$.

\clearpage
\newpage
\section{2DES doping dependence}
\begin{figure*}[h]
\includegraphics[width=14cm]{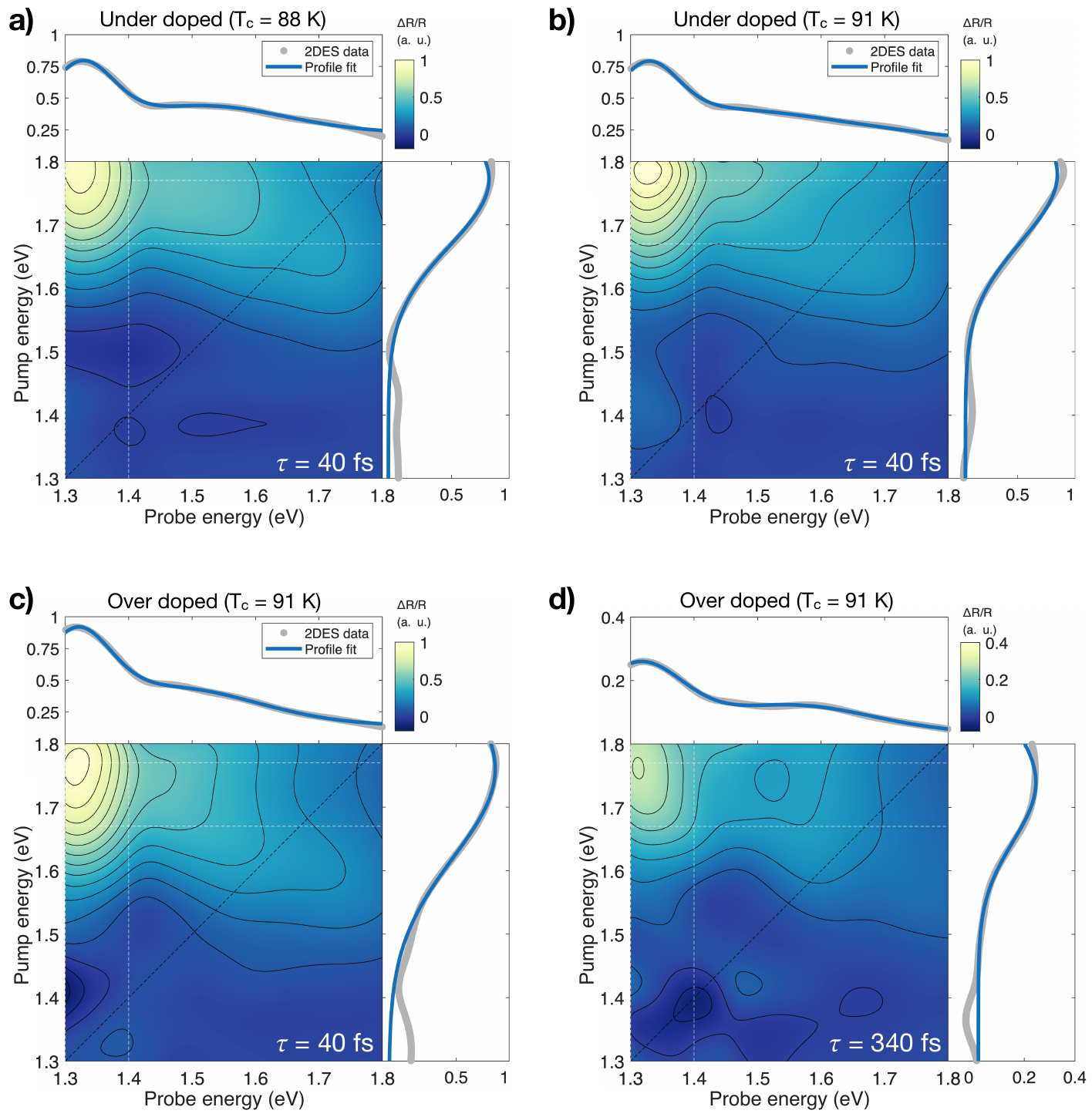}
\caption{2DES maps measured on different samples: (a) underdoped ($T_c = 88$~K, $p=0.13$), (b) underdoped ($T_c = 91$~K, $p=0.14$), and (c–d) overdoped ($T_c = 91$~K, $p=0.18$). Data were acquired at room temperature using a 400 \textmu\text{J/cm}$^2$ excitation fluence, with pump–probe time delays of $\tau = 40$~fs (a–c) and 340~fs (d).}
\label{fig:dopings maps}
\end{figure*}

Figure~\ref{fig:dopings maps} presents 2DES maps acquired at room temperature on Bi$_2$Sr$_2$Ca$_{0.92}$Y$_{0.08}$Cu$_2$O$_{8+\delta}$ samples spanning different doping levels: underdoped ($T_c = 88$~K and $T_c = 91$~K) and overdoped ($T_c = 91$~K). All maps were measured at an excitation fluence of $400$ \textmu\text{J/cm}$^2$. These data support the doping-dependent analysis presented in Fig.~6 of the main text. 

\clearpage
\bibliography{refs}